\documentclass[english]{article}
\usepackage[T1]{fontenc}
\usepackage[latin9]{inputenc}
\usepackage{geometry}
\usepackage{babel}
\usepackage{array}
\usepackage{textcomp}
\usepackage{url}
\usepackage{multirow}
\usepackage{amsmath}
\usepackage{amssymb}
\usepackage{graphicx}
\usepackage{setspace}
\usepackage[authoryear]{natbib}
\usepackage[unicode=true]
 {hyperref}

\makeatletter

\providecommand{\tabularnewline}{\\}

\usepackage{blkarray}

\makeatother

\begin{document}
\title{Log-density gradient covariance and automatic metric tensors for Riemann
manifold Monte Carlo methods\thanks{The author is wishes to express warm thanks to the Editor, Professor
Peltonen, an anonymous Associate Editor, two anonymous reviewers,
Nawaf Bou-Rabee, Roman Liesenfeld and Hans J. Skaug for comments and
suggestions for improvements on earlier versions of this paper. Kleppe
acknowledges support from Finansmarkedsfondet, grant \#337601.}}
\author{Tore Selland Kleppe\thanks{Department of Mathematics and Physics, University of Stavanger, Stavanger,
Norway. Email: tore.kleppe@uis.no}}
\maketitle
\begin{abstract}
A metric tensor for Riemann manifold Monte Carlo particularly suited
for non-linear Bayesian hierarchical models is proposed. The metric
tensor is built from symmetric positive semidefinite log-density gradient
covariance (LGC) matrices, which are also proposed and further explored
here. The LGCs generalize the Fisher information matrix by measuring
the joint information content and dependence structure of both a random
variable and the parameters of said variable. Consequently, positive
definite Fisher/LGC-based metric tensors may be constructed not only
from the observation likelihoods as is current practice, but also
from arbitrarily complicated non-linear prior/latent variable structures,
provided the LGC may be derived for each conditional distribution
used to construct said structures. The proposed methodology is highly
automatic and allows for exploitation of any sparsity associated with
the model in question. When implemented in conjunction with a Riemann
manifold variant of the recently proposed numerical generalized randomized
Hamiltonian Monte Carlo processes, the proposed methodology is highly
competitive, in particular for the more challenging target distributions
associated with Bayesian hierarchical models.
\end{abstract}

\section{Introduction}

Efficient posterior sampling for Bayesian statistical models has attracted
a substantial amount of research the last decades \citep[see e.g. ][]{2208.00646}.
Riemann manifold Monte Carlo methods \citep{girolami_calderhead_11}
are in particular well suited for posterior distributions exhibiting
complicated non-linear dependence structures and/or substantial differences
in scale across the target distribution. Posterior distributions with
these properties arise (among other) for Bayesian hierarchical models
which are widely used to model dependent data \citep[see e.g. ][]{doi:10.1080/10618600.2019.1584901}.
The successful application of the RMMC methods relies on the selection
of a suitable metric tensor, a symmetric, positive definite matrix-valued
function that should reflect the local scaling properties of the posterior
distribution in question. 

This article makes several contributions towards the end of selecting
metric tensors that are both of high quality and are easily applied
by non-experts in computational methods. The first contribution is
the introduction of the, to the author's knowledge, new concept Log-density
Gradient Covariance (LGC) and the development of some of its properties.
Informally, the LGC associated with some probability density, say
$\pi(\mathbf{x}|\boldsymbol{\theta})$ is defined to be the (necessarily
symmetric, positive semidefinite (SPSD)) covariance matrix of the
gradient of $\log\pi$ with respect to \emph{both} $\mathbf{x}$ and
$\boldsymbol{\theta}$. Consequently, the LGC generalizes the Fisher
information matrix \citep[see e.g. ][]{pawi:2001} (which is the covariance
matrix of the log-density gradient with respect to $\boldsymbol{\theta}$
only). Subject to regularity conditions, the LGC is equal to expected
negative Hessian of $\log\pi$ with respect to \emph{both} $\mathbf{x}$
and $\boldsymbol{\theta}$, and may informally speaking be used to
measure the information content and dependence structure between-
and among both $\mathbf{x}$ \emph{and} $\boldsymbol{\theta}$ in
cases where both $\mathbf{x}$ and $\boldsymbol{\theta}$ are sampled
(e.g. when $\mathbf{x}$ is a latent variable).

Secondly, a metric tensor is constructed from LGCs for a very broad
class of possibly non-linear models specified in terms of a sequence
of conditional distribution statements. Very few restrictions are
imposed, and in particular the class of models considered includes
non-linear hierarchical models, even with multiple- non-linearly coupled
layers of latent variables/priors. Consequently, guaranteed positive
definite Fisher/LGC-based metric tensors may be constructed not only
from the observation likelihoods as is current practice \citep{girolami_calderhead_11}.
Rather, a metric tensor may be constructed from arbitrarily complicated
non-linear prior/latent variable structures, provided the LGC may
be derived for each conditional distribution used to construct said
structures. The proposed metric tensor may be derived directly from
the model specification and does not involve any tuning parameters.
Third, an efficient and highly automatic numerical implementation
of said metric tensor based on Automatic Differentiation (AD) is proposed.
The implementation may exploit any sparsity of the metric tensor,
which for large scale hierarchical models is essential in a performance
perspective.

The proposed metric tensor could in principle be used in conjunction
with any Riemann manifold Monte Carlo method, e.g. Riemann manifold
Hamiltonian Monte Carlo or Riemann manifold Langevin dynamics \citep{girolami_calderhead_11}.
However, in this article, the illustrations are done based on a Riemann
manifold variant of the numerical generalized randomized Hamiltonian
Monte Carlo (NGRHMC) method of \citep{kleppe_CTHMC}. NGRHMC processes
are continuous time piecewise deterministic processes \citep[see e.g. ][]{fearnhead2018}
with Hamiltonian deterministic dynamics \citep{bou-rabee2017} which
are implemented using adaptive numerical ordinary differential equations
(ODEs) solvers. The usage of such ODE solvers introduces small biases,
but at the same time avoids computationally intensive- and difficult
to tune implicit symplectic integrators commonly used in Riemann manifold
Hamiltonian Monte Carlo. The application of NGRHMC allows a clean
comparison between samplers based on the proposed Riemann manifold
Hamiltonian dynamics and conventional Euclidean metric Hamiltonian
dynamics, as the same numerical ODE solver may be used in both cases. 

Finally, the paper contains several numerical illustrations, which
benchmarks the proposed methodology against relevant alternatives.
It is demonstrated that the proposed methodology may lead to substantial
speed-ups in sampling efficiency (or expand the set of target distributions
that may reliably sampled using HMC-like methods without introducing
complicated rescaling methodology), in particular for challenging
target distributions associated with large Bayesian hierarchical models. 

Below, Section \ref{sec:Background} provides background material
and relation to literature, and Section \ref{sec:Log-density-gradient-covariance}
introduces the LGC and discusses some of its properties. Section \ref{sec:Metric-tensors-based}
derives a metric tensor based on LGC provides some illustrations of
the properties of the metric tensor and discusses automatic implementation.
Numerical examples and benchmarking are found in Sections \ref{sec:Examples}
and \ref{sec:A-Wishart-transition_SV}, and Section \ref{sec:Discussion}
provides discussion. The article is accompanied by an online appendix
which provides proofs and additional information in several regards.

\section{Background\label{sec:Background}}

This section provides necessary background and fixes notation. For
the purpose of readability, the notation and language is as far as
possible avoiding differential-geometric nomenclature. Further, the
paper assumes familiarity with Markov chain Monte Carlo (MCMC) methods
and in particular Hamiltonian Monte Carlo (HMC) methods, for which
e.g. \citet{1206.1901,girolami_calderhead_11,bou-rabee_sanz-serna_2018}
may serve as references.

The paper considers a continuous target density $\mathbf{\pi}(\mathbf{q})$
with density kernel $\bar{\pi}(\mathbf{q})\propto\pi(\mathbf{q})$,
$\mathbf{q}\in\mathbb{R}^{D}$ (with respect to the Euclidean geometry)
that allows evaluation. In the following, $\mathcal{N}(\mathbf{x}|\boldsymbol{\mu},\boldsymbol{\Sigma})$
denotes the density of a $N(\boldsymbol{\mu},\boldsymbol{\Sigma})$
random vector evaluated at $\mathbf{x}$. $\mathbf{0}_{d}\in\mathbb{R}^{d}$
and $\mathbf{0}_{d,n}\in\mathbb{R}^{d\times n}$ denote vectors and
matrices of only zeros. For stacking of two vectors, say $\mathbf{x}\in\mathbb{R}^{d}$
and $\mathbf{y\in\mathbb{R}}^{n}$ into $\mathbf{z}=[\mathbf{x}^{T}\quad\mathbf{y}^{T}]^{T}\in\mathbb{R}^{d+n}$,
the shorthand notation $(\mathbf{x},\mathbf{y})$ is sometimes used.
Further, $\nabla_{\mathbf{x}}f(\mathbf{x})\in\mathbb{R}^{d}$ denotes
the gradient of $f:\mathbb{R}^{d}\mapsto\mathbb{R}$, and $\nabla_{\mathbf{x}}^{2}f(\mathbf{x})\in\mathbb{R}^{d\times d}$
the Hessian of $f$. Finally, $\nabla_{\mathbf{x}}g(\mathbf{x})\in\mathbb{R}^{p\times d}$
denotes the Jacobian of $g:\mathbb{R}^{d}\mapsto\mathbb{R}^{p}$.

\subsection{Metric tensors and Riemann manifold Hamiltonian dynamics}

Broadly speaking, Riemann manifold MCMC methods \citep{girolami_calderhead_11}
rely on defining the proposal mechanism of the MCMC method on a (non-trivial)
Riemann manifold rather than the conventional Euclidean space $\mathbb{R}^{D}$.
The Riemann manifold under consideration here may be characterized
in terms of the \emph{metric tensor} $\mathbf{G}(\mathbf{q})\in\mathbb{R}^{D\times D}$,
a smooth symmetric, positive definite (SPD) matrix-valued function
for each $\mathbf{q}\in\mathbb{R}^{D}$. For purposes of this paper,
it suffices to think of the metric tensor as giving the distance between
two infinitesimally separated points $\mathbf{q}\in\mathbb{R}^{D}$
and $\mathbf{q}+\boldsymbol{\delta}\mathbf{q}\in\mathbb{R}^{D}$ to
be $\sqrt{(\boldsymbol{\delta}\mathbf{q})^{T}\mathbf{G}(\mathbf{q})\boldsymbol{\delta}\mathbf{q}}$
(rather than the conventional Euclidean distance $\sqrt{(\boldsymbol{\delta}\mathbf{q})^{T}\boldsymbol{\delta}\mathbf{q}}$). 

To leverage the flexibility afforded by introducing a non-trivial
Riemann manifold for constructing HMC-like RMMC methods targeting
$\pi$, \citet{girolami_calderhead_11} suggested using the dynamics
associated with the Hamiltonian 
\begin{equation}
\mathcal{H}(\mathbf{q},\mathbf{p})=-\log\bar{\pi}(\mathbf{q})+\frac{1}{2}\log|\mathbf{G}(\mathbf{q})|+\frac{1}{2}\mathbf{p}^{T}\left[\mathbf{G}(\mathbf{q})\right]^{-1}\mathbf{p}\label{eq:hamiltonian}
\end{equation}
as the proposal mechanism. Here $\mathbf{p}\in\mathbb{R}^{D}$ is
the fictitious momentum variable. The dynamics associated with (\ref{eq:hamiltonian})
are governed by Hamilton's equations, which amounts to
\begin{align}
\dot{\mathbf{q}}(t) & =\nabla_{\mathbf{p}}\mathcal{H}(\mathbf{q}(t),\mathbf{p}(t))=\mathbf{G}(\mathbf{q}(t))^{-1}\mathbf{p}(t),\label{eq:ham_eq_1}\\
\dot{\mathbf{p}}(t) & =-\nabla_{\mathbf{q}}\mathcal{H}(\mathbf{q}(t),\mathbf{p}(t)).\label{eq:ham_eq_2}
\end{align}
The Boltzmann-Gibbs (BG) distribution for $\mathbf{z}=(\mathbf{q}^{T},\mathbf{p}^{T})^{T}$,
associated with (\ref{eq:hamiltonian}) is given by 
\[
\pi(\mathbf{z})=\pi(\mathbf{q},\mathbf{p})=\pi(\mathbf{q})\mathcal{N}(\mathbf{p}|\mathbf{0}_{D},\mathbf{G}(\mathbf{q})).
\]
The dynamics (\ref{eq:ham_eq_1},\ref{eq:ham_eq_2}) preserve both
the Hamiltonian (i.e. total energy), and are also volume preserving.
Consequently, (\ref{eq:ham_eq_1},\ref{eq:ham_eq_2}) preserve the
BG distribution in the sense that for any initial configuration $\mathbf{z}(0)\sim\pi(\mathbf{z})$,
then $\mathbf{z}(T)\sim\pi(\mathbf{z})$ for any $T>0$ (provided
$\mathbf{z}(t)$ solves Hamilton's equations (\ref{eq:ham_eq_1},\ref{eq:ham_eq_2})
for each $t\in[0,T]$). Clearly, the original target $\pi(\mathbf{q})$
is the $\mathbf{q}$-marginal of the BG-distribution. 

\subsection{Riemann manifold HMC}

Arguably, the most promising general purpose RMMC method is Riemann
manifold HMC (RMHMC) \citep{girolami_calderhead_11}. RMHMC is most
easily explained as a discrete time MCMC algorithm targeting $\pi(\mathbf{z})$
(and samples targeting $\pi(\mathbf{q})$ may subsequently be obtained
by discarding the $\mathbf{p}$-coordinates of samples targeting $\pi(\mathbf{z})$).
Each transition of RMHMC, say from $\mathbf{z}_{(i)}=(\mathbf{q}_{(i)},\mathbf{p}_{(i)})$
to $\mathbf{z}_{(i+1)}$ involves two steps, where the first step
is updating the momentum $\mathbf{p}_{(i)}^{*}\sim\pi(\mathbf{p}|\mathbf{q}_{(i)})=\mathcal{N}(\mathbf{p}|\mathbf{0}_{D},\mathbf{G}(\mathbf{q}_{(i)}))$.
In the second step, a fixed/random number of numerical time-integration
steps applied to the ODE (\ref{eq:ham_eq_1},\ref{eq:ham_eq_2}) with
initial configuration $(\mathbf{q}_{(i)},\mathbf{p}_{(i)}^{*})$ are
computed. The final state of the time-integration process is either
accepted or rejected as $\mathbf{z}_{(i+1)}$ according to a Metropolis-Hastings
(MH) mechanism in order to adjust for errors introduced by the numerical
integration relative to the exact solution of (\ref{eq:ham_eq_1},\ref{eq:ham_eq_2}). 

Provided the time-integration is done using a symplectic/time-reversible
method \citep[see e.g. ][]{sanzSerna_Calvo,Leimkuhler:2004}, the
accept probability of the MH step takes a particularly simple form.
However, symplectic numerical methods for (\ref{eq:ham_eq_1},\ref{eq:ham_eq_2})
are necessarily implicit, with each integration step requiring the
iterative solution of a set of $D$ non-linear equations involving
the $\mathbf{q}$-gradient of the Hamiltonian (\ref{eq:hamiltonian}).
Further, to ensure stability and convergence of these iterative processes,
it is typically necessary to use very short/many time-integration
steps in each transition. Consequently, unless special structures
in the model may be exploited \citep[see e.g. ][]{NIPS2014_5591,doi:10.1080/10618600.2019.1584901},
RMHMC may be very computationally demanding in practice for general
model. It is also worth mentioning that MH adjustment mechanisms for
RMHMC-like methods may be implemented with explicit time-reversible
(but not volume preserving/symplectic) integrators, but this in general
leads to highly nontrivial calculations for the accept probability
\citep[see e.g. ][]{doi:10.1080/10618600.2014.902764}.

\subsection{Riemann manifold Numerical Generalized Randomized HMC processes}

Rather than conventional discrete time RMHMC methods, the illustrations
of the proposed metric tensor in this paper are done using Riemann
manifold variants of numerical Generalized Randomized HMC (GRHMC)
processes \citep{kleppe_CTHMC}. The Riemann manifold variant of numerical
GRHMC processes \citep[see][supplementary material]{kleppe_CTHMC},
implemented with general purpose explicit adaptive first order ODE
solvers \citep[see e.g. ][]{10.5555/153158}, bypasses the need for
iterative non-linear equation solving, while still exploiting the
conservative nature (\ref{eq:ham_eq_1},\ref{eq:ham_eq_2}) and the
freedom to choose $\mathbf{G}(\mathbf{q})$. The savings in computing
time comes at the cost of arbitrarily small errors introduced by the
non-symplectic and un-adjusted ODE solver. 

An added benefit of considering numerical GRHMC processes rather than
RMHMC is that a clean comparison between Riemann manifold-based methods
based on the proposed metric tensor, and Euclidean metric (i.e., $\mathbf{G}(\mathbf{q})=\mathbf{M}$
for some fixed SPD mass matrix $\mathbf{M}$) numerical GRHMC methods
may be carried out without having to take into account the effects
of difficult to tune/computationally costly symplectic integrators.
Rather, the same adaptive time step integrator, with the same error
tolerances may be used both for Riemann manifold- and fixed metric
methods.

Riemann manifold GRHMC processes, say $\mathbf{Z}(t)=(\mathbf{Q}(t),\mathbf{P}(t)),\;t\geq0$,
are continuous time processes that may be specified so that $\mathbf{Q}(t)$
has an arbitrary continuous stationary distribution. The processes
are special cases of piecewise deterministic Markov processes \citep{davis_PDP_book,fearnhead2018,1707.05296}.
For simplicity, in this paper, only constant event rate processes,
where events occur according to a time-homogenous Poisson process
with intensity $\lambda$ are considered. Between events, $\mathbf{Z}(t)$
solves (\ref{eq:ham_eq_1},\ref{eq:ham_eq_2}). At events times $t$,
the momentum coordinate $\mathbf{P}(t)$ is updated according to $\pi(\mathbf{p}|\mathbf{q}=\mathbf{Q}(t))$.
Continuous time trajectories are simulated for a pre-specified time
interval $[0,T_{\max}]$, and position coordinate is subsequently
sampled at discrete times, say $\left\{ \mathbf{q}_{i}=\mathbf{Q}(i\Delta)\right\} _{i}$
for some suitable time increment $\Delta$. The discrete time samples
$\left\{ \mathbf{q}_{i}\right\} _{i}$ may be used in the same manner
as samples from conventional (discrete time) MCMC methods. 

The (Riemann manifold and Euclidean metric) numerical GRHMC processes
used for illustration are implemented in the pdmphmc R-package (\url{https://github.com/torekleppe/pdmphmc}).
For improved numerical performance, the simulations are done in standardized
variables (see Appendix \ref{sec:Details-of-the-numerical} for details
on standardization and other aspects related to the numerical implementation).
This standardization has the added benefit of making the unit of process
time $t$ comparable across many models/manifolds, and as a rule of
thumb (obtained by trial and error), one should expect on the order
of 0.2 effective samples per unit of process time. 

\subsection{Metric tensors in literature}

So far, the metric tensor $\mathbf{G}(\mathbf{q})$ has been left
unspecified. The overarching aim of working with a non-constant $\mathbf{G}(\mathbf{q})$
is to ensure that the dynamics (\ref{eq:ham_eq_1},\ref{eq:ham_eq_2})
result in efficient exploration of the target distribution. In most
applications of Riemann manifold HMC methods, $\mathbf{G}(\mathbf{q})$
is chosen to be some sort of positive definite approximation to the
negative Hessian of the log-target density, i.e. $-\nabla_{\mathbf{q},}^{2}\log\pi(\mathbf{q})$.
Such choices may be motivated by that in the flow of $\mathbf{q}$
associated with (\ref{eq:ham_eq_1},\ref{eq:ham_eq_2}), the log-target
gradient gets scaled by $\mathbf{G}^{-1}(\mathbf{q})$ \citep[see e.g. ][Equation 9]{Kleppe2017}.
It is well known from the numerical optimization literature \citep[see e.g. ][]{noce:wrig:1999}
that scaling the target function gradient using some form of positive
definite approximation to the inverse target Hessian typically result
in moves well adapted to the target distribution. 

When the target distribution is the posterior distribution of a \emph{non-hierarchical}
statistical model, i.e. $\pi(\mathbf{q})\propto\pi(\mathbf{y}|\mathbf{q})p(\mathbf{q})$
(where $\pi(\mathbf{y}|\mathbf{q})$ is the likelihood function for
observations $\mathbf{y}$ and parameters $\mathbf{q}$, and $p(\mathbf{q})$
is the prior.), \citet{girolami_calderhead_11} suggest using the
metric tensor 
\begin{equation}
\mathbf{G}(\mathbf{q})=\mathcal{F}(\mathbf{q})+\mathcal{R}.\label{eq:GC_MT}
\end{equation}
Here $\mathcal{F}(\mathbf{q})=-E_{\mathbf{y}|\mathbf{q}}[\nabla_{\mathbf{q}}^{2}\log\pi(\mathbf{y}|\mathbf{q})]$
is the Fisher information matrix \citep[see e.g. ][]{pawi:2001} associated
with the likelihood function, and $\mathcal{R}=-\nabla_{\mathbf{q}}^{2}\log p(\hat{\mathbf{q}}),\;\hat{\mathbf{q}}=\arg\max_{\mathbf{q}}p(\mathbf{q})$,
is the negative Hessian at the maximizer of the prior, a common approximation
to the precision matrix of $p(\mathbf{q})$ \citep[see e.g. ][ for a discussion of such Hessian-based approximations to precision matrices]{GelmanBDA3}.
Based on that $\mathcal{F}(\mathbf{q})$ is the natural metric tensor
for the parameter space Riemann manifold associated with the statistical
model $\pi(\mathbf{y}|\mathbf{q})$, \citet[ Section 4]{girolami_calderhead_11}
provide a discussion of why (\ref{eq:GC_MT}) constitutes a suitable
metric tensor. See also \citet{10.1162/089976698300017746} for further
discussion of the application of Fisher information for the closely
related natural gradient in non-hierarchical models.

\citet{1212.4693,Kleppe2017} propose to use positive definite approximations
to/modifications of $-\nabla_{\mathbf{q}}^{2}\log\pi(\mathbf{q})$
as the metric tensor. Such procedures have the benefit of allowing
for a high degree of automation, as $-\nabla_{\mathbf{q}}^{2}\log\pi(\mathbf{q})$
may be computed from a program specifying the log-target density using
automatic differentiation (AD) techniques \citep{grie:2000}. \citet{1212.4693}
uses a full eigen-decomposition and modifies any small positive or
negative eigenvalues of the negative Hessian. \citet{Kleppe2017}
on the other hand uses modified Cholesky factorization that exploits
any sparsity of the Hessian, commonly present under hierarchical models
\citep[see e.g. ][]{RSSB:RSSB700}, to a similar end. Common for both
techniques is that they require the non-trivial selection of a regularization
parameter which chooses a tradeoff between the smoothness of the resulting
$\mathbf{G}(\mathbf{q})$ against the difference between $-\nabla_{\mathbf{q}}^{2}\log\pi(\mathbf{q})$
and $\mathbf{G}(\mathbf{q})$. Further, computing the required derivatives
of the Hamiltonian (\ref{eq:ham_eq_2}) effectively amounts to third
order AD, which may both be computationally demanding and require
highly specialized techniques or additional input by the user if sparsity
is to be exploited.

Recently, \citet{2202.00755} proposed the Monge metric, which in
the present notation amounts to $\mathbf{G}(\mathbf{q})=\mathbf{I}_{D}+\alpha^{2}[\nabla_{\mathbf{q}}\log\pi(\mathbf{q})][\nabla_{\mathbf{q}}\log\pi(\mathbf{q})]^{T}$
for with $\alpha$ being a tuning parameter. The Monge metric also
does not assume any particular structure on the model, and would allow
implementation based on second order AD. The identity plus rank 1
update structure of the Monge metric affords substantial savings in
the numerical linear algebra involved in each update, but it is not
clear how to choose $\alpha$ for any given statistical model. Note
also that the expectation of the gradient outer product of the Monge
metric is the Fisher information provided a similar model structure
as for (\ref{eq:GC_MT}) and flat priors.

To reduce the cost of each RMHMC update for hierarchical models, certain
structure can be imposed on the metric tensor. \citet{NIPS2014_5591}
proposes semi-separable HMC, and the dynamic rescaling method of \citet{doi:10.1080/10618600.2019.1584901}
may also be interpreted in terms a metric tensor with certain properties
which would simplify RMHMC sampling. Both approaches are based on
Fisher information matrices, but requires different, rather strict
assumptions on the model which does not lend themselves easily to
automatic implementation.

In what follows, a new metric tensor, along with an efficient and
automatic method of computation of this metric tensor is proposed.
The proposed metric tensor may be seen as a generalization of the
Fisher-based metric (\ref{eq:GC_MT}) of \citet{girolami_calderhead_11}
to hierarchical/latent variable models that allows for a high degree
of automation.

\section{Log-density gradient covariance\label{sec:Log-density-gradient-covariance}}

Before discussing metric tensors per se, the log-density gradient
covariance (LGC) is introduced. The LGC generalizes the Fisher information
matrix \citep[see e.g. ][]{pawi:2001} for sufficiently smooth probability
densities, and will constitute an important building block for the
proposed metric tensor. 

\subsection{Log-density gradient covariance}

\textbf{Assumption 1:} \emph{Probability density $\pi(\mathbf{x}|\boldsymbol{\theta})$
on $\mathbb{R}^{d}$ has continuous first order derivatives w.r.t.
$\mathbf{x}$ for each $\boldsymbol{\theta}\in\Omega\subseteq\mathbb{R}^{p}$
where $\Omega$ is the set of allowed parameters. }

Under Assumption 1, the LGC $\mathbb{V}_{\pi}[\mathbf{x}|\boldsymbol{\theta}]$
associated with probability density $\pi(\mathbf{x}|\boldsymbol{\theta})$
is defined as
\[
\mathbb{V}_{\pi}[\mathbf{x}|\boldsymbol{\theta}]=\underset{\pi(\mathbf{x}|\boldsymbol{\theta})}{Var}\left[\nabla_{(\mathbf{x},\boldsymbol{\theta})}\log\pi(\mathbf{x}|\boldsymbol{\theta})\right]=\left[\begin{array}{cc}
\mathbf{\mathcal{V}}(\boldsymbol{\theta}) & \mathcal{W}(\boldsymbol{\theta})\\
\mathcal{W}^{T}(\boldsymbol{\theta}) & \mathcal{F}(\boldsymbol{\theta})
\end{array}\right],
\]
where the blocks $\mathbf{\mathcal{V}}(\boldsymbol{\theta})\in\mathbb{R}^{d\times d}$,
$\mathbf{\mathcal{W}}(\boldsymbol{\theta})\in\mathbb{R}^{d\times p}$
and $\mathcal{F}(\boldsymbol{\theta})\in\mathbb{R}^{p\times p}$ conform
in sizes with the sizes of $\mathbf{x}$ and $\boldsymbol{\theta}$.
Clearly $\mathcal{F}(\boldsymbol{\theta})$ is the Fisher information
matrix associated with $\pi(\mathbf{x}|\boldsymbol{\theta})$. Being
proper covariance matrices, both $\mathbb{V}_{\pi}[\mathbf{x}|\boldsymbol{\theta}]$
and $\mathbf{\mathcal{V}}(\boldsymbol{\theta})$ (in addition to $\mathcal{F}(\boldsymbol{\theta})$
obviously) are symmetric and positive semi-definite. In general, the
LGC is a SPSD matrix-valued function of $\boldsymbol{\theta}$, and
sometimes the notation $\mathbb{V}_{\pi}[\mathbf{x}|\boldsymbol{\theta}](\boldsymbol{\theta})$
is needed. 

\subsection{Basic properties of the log-density gradient and the LGC}

The log-density gradient, $\nabla_{\mathbf{(\mathbf{x},\boldsymbol{\theta})}}\log\pi(\mathbf{x}|\boldsymbol{\theta})$,
and the LGC have properties that mirror those of the score function
and Fisher information:

\textbf{Proposition 1}: \emph{Under Assumption 1,
\[
I:\;\underset{\pi(\mathbf{x}|\boldsymbol{\theta})}{E}\left[\nabla_{\mathbf{(\mathbf{x},\boldsymbol{\theta})}}\log\pi(\mathbf{x}|\boldsymbol{\theta})\right]=\mathbf{0}_{d+p},
\]
and}
\[
II:\;\mathbb{V}_{\pi}[\mathbf{x}|\boldsymbol{\theta}]=-\underset{\pi(\mathbf{x}|\boldsymbol{\theta})}{E}\left[\nabla_{\mathbf{\mathbf{(\mathbf{x},\boldsymbol{\theta})}}}^{2}\log\pi(\mathbf{x}|\boldsymbol{\theta})\right].
\]
The proof is provided in Appendix \ref{subsec:Proof-of-Proposition1}. 

The second part of the proposition indicates, via the established
explicit relation to the Hessian matrix with respect to $(\mathbf{x},\boldsymbol{\theta})$,
that $\mathbb{V}_{\pi}[\mathbf{x}|\boldsymbol{\theta}]$ is a sensible
``scale matrix'' for statistical computing purposes in cases where
variation in $\mathbf{x},\boldsymbol{\theta}$ jointly is considered. 

The above proposition relies critically on the smoothness of Assumption
1, which in turn implies that $\int\frac{\partial}{\partial x_{i}}\pi(\mathbf{x}|\boldsymbol{\theta)}d\mathbf{x}=\lim_{x_{i}\rightarrow\infty}\pi(\mathbf{x}|\boldsymbol{\theta)}-\lim_{x_{i}\rightarrow-\infty}\pi(\mathbf{x}|\boldsymbol{\theta)}=0$
for all $i=1,\dots,d$. Failures to be sufficiently smooth, e.g. the
exponential distribution (interpreted as a distribution on $\mathbb{R}$
with density evaluating to 0 for negative arguments), may in certain
cases be worked around by transformations of $\mathbf{x}$, see e.g.
ExpGamma distribution below. The regular Gamma distribution is sufficiently
smooth for shape parameter $>2$ as then it will have continuous first
order derivative with respect to $\mathbf{x}$ everywhere.

\subsection{A transformation result}

It is well known that the Fisher information matrix for some alternative
parameter, say $\boldsymbol{\eta}$, may be expressed in terms the
Fisher information associated with the original parameter, say $\boldsymbol{\theta}=\boldsymbol{\Psi}(\boldsymbol{\eta})$.
A similar result can be derived for the LGC subject to transformations
between $(\mathbf{x},\boldsymbol{\theta})$ and $(\mathbf{z},\boldsymbol{\eta})$
of the form
\begin{equation}
\mathbf{x}=\mathbf{a}(\boldsymbol{\eta})+\mathbf{B}\mathbf{z},\;\boldsymbol{\theta}=\boldsymbol{\Psi}(\boldsymbol{\eta}),\label{eq:repar}
\end{equation}
where it is assumed that matrix $\mathbf{B}$ is invertible (and hence
the dimensions of $\mathbf{x}$ and $\mathbf{z}$ are equal). Denote
by $p(\mathbf{z}|\boldsymbol{\eta})=\pi(\mathbf{a}(\boldsymbol{\eta})+\mathbf{B}\mathbf{z}|\boldsymbol{\Psi}(\boldsymbol{\eta}))|\mathbf{B}|$
the density of $\mathbf{z}|\boldsymbol{\eta}$ implied by $\mathbf{x}|\boldsymbol{\theta}$
being distributed according to $\pi(\mathbf{x}|\boldsymbol{\theta})$
and (\ref{eq:repar}). Then the LGC associated with $p(\mathbf{z}|\boldsymbol{\eta})$
may be expressed in terms of $\mathbb{V}_{\pi}[\mathbf{x}|\boldsymbol{\theta}]$,
namely
\begin{equation}
\mathbb{V}_{p}[\mathbf{z}|\boldsymbol{\eta}]=\mathbf{U}(\boldsymbol{\eta})^{T}\left\{ \mathbb{V}_{\pi}[\mathbf{x}|\boldsymbol{\theta}](\boldsymbol{\Psi}(\boldsymbol{\eta}))\right\} \mathbf{U}(\boldsymbol{\eta}),\;\mathbf{U}(\boldsymbol{\eta})=\nabla_{(\mathbf{z},\boldsymbol{\eta})}(\mathbf{x},\boldsymbol{\theta})=\left[\begin{array}{cc}
\mathbf{B} & \nabla_{\boldsymbol{\eta}}\mathbf{a}(\boldsymbol{\eta})\\
\mathbf{0} & \nabla_{\boldsymbol{\eta}}\boldsymbol{\Psi}(\boldsymbol{\eta})
\end{array}\right].\label{eq:repar_LCG}
\end{equation}
The rather elementary proof of (\ref{eq:repar_LCG}) is detailed in
Appendix \ref{subsec:Proof-of-()}. Clearly, setting $\mathbf{a}(\boldsymbol{\eta})=\mathbf{0},\;\mathbf{B}=\mathbf{I}$
recovers the conventional re-parameterization formula for the Fisher
information \citep{pawi:2001} (with the cross-information modified
to be $\mathcal{W}[\mathbf{z}|\boldsymbol{\eta}]=\mathcal{W}[\mathbf{x}|\boldsymbol{\theta}=\boldsymbol{\Psi}(\boldsymbol{\eta})]\nabla_{\boldsymbol{\eta}}\boldsymbol{\Psi}(\boldsymbol{\eta})$).
Further, the LGC exhibit intuitive behavior by being unchanged under
constant (w.r.t. parameters) location shifts of the random variable
($\mathbf{B}=\mathbf{I}$, $\boldsymbol{\theta}=\boldsymbol{\eta}$
and $\nabla_{\boldsymbol{\eta}}\mathbf{a}=\mathbf{0}$ so that $\mathbf{U}=\mathbf{I}$).
Even further, (\ref{eq:repar_LCG}) entails that the LGC random variable
block $\mathcal{V}$ scales as conventional precision matrix under
invertible linear transformations of the random variable. 

\subsection{Examples of LGCs}

This section gives some examples of LGCs for common probability distributions.
The Gaussian distribution with density $\mathcal{N}(x|\mu,\sigma^{2})$
has the LGC
\begin{equation}
\mathbb{V}_{\mathcal{N}}[x|(\mu,\sigma)]=\sigma^{-2}\left[\begin{array}{ccc}
1 & -1 & 0\\
-1 & 1 & 0\\
0 & 0 & 2
\end{array}\right]\label{eq:1dnormal_LGC}
\end{equation}
More generally, for a multivariate Gaussian distribution, say $\mathcal{N}(\mathbf{x}|\boldsymbol{\mu},\mathbf{P}^{-1}(\boldsymbol{\omega}))$
where the precision matrix $\mathbf{P}(\boldsymbol{\omega})$ depends
on a parameter vector $\boldsymbol{\omega}$, it is clear that
\[
\mathbb{V}_{\mathcal{N}}[\mathbf{x}|(\boldsymbol{\mu},\boldsymbol{\omega})]=\left[\begin{array}{ccc}
\mathbf{P}(\boldsymbol{\omega}) & -\mathbf{P}(\boldsymbol{\omega}) & \mathbf{0}\\
-\mathbf{P}(\boldsymbol{\omega}) & \mathbf{P}(\boldsymbol{\omega}) & \mathbf{0}\\
\mathbf{0} & \mathbf{0} & \mathcal{F}_{\boldsymbol{\omega}}
\end{array}\right]
\]
where $\mathcal{F}_{\boldsymbol{\omega}}$ is the Fisher information
of $\mathcal{N}(\mathbf{x}|\boldsymbol{\mu},\mathbf{P}^{-1}(\boldsymbol{\omega}))$
with respect to $\boldsymbol{\omega}$. 

For densities $\pi(\mathbf{y}|\boldsymbol{\theta})$ that do not have
a everywhere continuous derivative with respect $\mathbf{x}$, LGCs
may be derived after first transforming $\mathbf{y}$. Examples include
the ExpGamma-distribution (named analogous with the LogNormal distribution),
i.e. if $X\sim ExpGamma(\alpha,\beta)$, then $Y=\exp(X)\sim Gamma(\alpha,\beta)$
where $\beta$ is the scale parameter. The ExpGamma distribution has
density $\pi(x|\alpha,\beta)\propto\exp(\alpha x-\beta^{-1}\exp(x)),\;x\in\mathbb{R}$,
and yields the LGC
\begin{equation}
\mathbb{V}_{\pi}[x|(\alpha,\beta)]=\left[\begin{array}{ccc}
\alpha & -1 & -\alpha\beta^{-1}\\
-1 & \Psi^{\prime}(\alpha) & \beta^{-1}\\
-\alpha\beta^{-1} & \beta^{-1} & \alpha\beta^{-1}
\end{array}\right]\label{eq:expGammaLGC}
\end{equation}
where $\Psi^{\prime}(a)=\frac{d^{2}}{da^{2}}\log(\Gamma(a))$. 

Another such example would be the InverseLogitBeta, defined via $X\sim InverseLogitBeta(a,b)\Rightarrow Y=\text{logit}^{-1}(X)\sim\text{Beta}(a,b)$,
where $\text{logit}^{-1}(x)=\frac{\exp(x)}{\exp(x)+1}$. The InverseLogitBeta
distribution, which has density $\pi(x|a,b)\propto\left[\exp(x)/(1+\exp(x))\right]^{a}\left[1/(1+\exp(x))\right]^{b},\;x\in\mathbb{R}$,
has LGC given by 
\[
\mathbb{V}_{\pi}[x|(a,b)]=\left[\begin{array}{ccc}
\frac{ab}{a+b+1} & -\frac{b}{a+b} & \frac{a}{a+1}\\
-\frac{b}{a+b} & \Psi^{\prime}(a)-\Psi^{\prime}(a+b) & -\Psi^{\prime}(a+b)\\
\frac{a}{a+1} & -\Psi^{\prime}(a+b) & \Psi^{\prime}(b)-\Psi^{\prime}(a+b)
\end{array}\right]
\]
Note that in the context of statistical computing using HMC-like methods,
it is common practice to transform constrained variables into un-constrained
ones (to obtain continuous first order derivatives), as in the two
latter examples, before sampling is performed. E.g. Stan also uses
internally the $\log$- and $\text{logit}$-transforms to arrive at
unconstrained variables from lower-bounded and compactly supported
variables respectively \citep{JSSv076i01}. Hence, for application
of the LGC within statistical computing, the requirement that the
involved densities fulfill Assumption 1 is not too restrictive.

\section{Metric tensors based on LGC\label{sec:Metric-tensors-based}}

Commonly, Bayesian hierarchical models are built from sequences of
known conditional distributions, for which deriving LGCs (or Fisher
information matrices in the case of discrete observation likelihoods)
is usually relatively easy. This section discusses how leverage such
LGCs to arrive at the proposed metric tensor for models built from
such sequences of conditional distributions with potentially non-linear
interconnections which may generate complicated dependence structures.

\subsection{Model formulation and notation}

The proposed methodology assumes that:\textbf{}\\
\textbf{Assumption 2:} \emph{The joint posterior distribution of statistical
model under consideration may be written as} 
\begin{equation}
\log\pi(\mathbf{q})\propto\sum_{\ell}\log\pi_{\ell}(\boldsymbol{\phi}_{\ell}(\mathbf{q})|\mathbf{\boldsymbol{\psi}}_{\ell}(\mathbf{q}))\label{eq:target_rep}
\end{equation}
\emph{for suitably chosen probability densities/mass functions $\{\pi_{\ell}\}_{\ell}$,
``argument functions'' $\{\boldsymbol{\phi}_{\ell}(\mathbf{q})\}_{\ell}$
and ``parameter functions'' $\{\boldsymbol{\mathbf{\psi}}_{\ell}(\mathbf{q})\}_{\ell}$}.
\textbf{}\\
\textbf{Assumption 3:} \emph{Provided $\nabla_{\mathbf{q}}\boldsymbol{\phi}_{\ell}(\mathbf{q})\neq\mathbf{0}$,
then $\pi_{\ell}(\mathbf{x}_{\ell}|\boldsymbol{\theta}_{\ell})$ admit
a LGC $\mathbb{V}_{\pi_{\ell}}[\mathbf{x}_{\ell}|\boldsymbol{\theta}_{\ell}]$.
If $\nabla_{\mathbf{q}}\boldsymbol{\phi}_{\ell}(\mathbf{q})=\mathbf{0}$,
then $\pi_{\ell}(\mathbf{x}_{\ell}|\boldsymbol{\theta}_{\ell})$ admits
a Fisher information matrix}.\\
Note that both $\boldsymbol{\mathbf{\psi}}_{\ell}(\mathbf{q})$ and
$\boldsymbol{\phi}_{\ell}(\mathbf{q})$ may be constant with respect
to the sampled quantity $\mathbf{q}$. E.g., $\boldsymbol{\psi}_{\ell}$
being some fixed hyper-parameters if $\pi_{\ell}$ is a prior, or
$\boldsymbol{\phi}_{\ell}(\mathbf{q})$ being equal to a set of observations/data.
In cases where $\nabla_{\mathbf{q}}\boldsymbol{\phi}_{\ell}(\mathbf{q})=\mathbf{0}$
and $\pi_{\ell}$ does not admit a LGC (e.g. a discrete distribution),
$\mathcal{V}_{\ell}$ and $\mathcal{W}_{\ell}$ are taken to be the
zero-matrices in the subsequent derivations.

\subsection{Proposed metric tensor\label{subsec:Proposed-metric-tensor}}

Based on the above model formulation and assumptions 2 and 3, this
Section proposes a metric tensor suitable for statistical computing
applications. Define the Jacobian matrix
\[
\mathbf{J}_{\ell}(\mathbf{q})=\left[\begin{array}{c}
\nabla_{\mathbf{q}}\boldsymbol{\phi}_{\ell}(\mathbf{q})\\
\nabla_{\mathbf{q}}\boldsymbol{\psi}_{\ell}(\mathbf{q})
\end{array}\right].
\]
In the current paper, it is proposed to use 
\begin{equation}
\mathbf{G}(\mathbf{q})=\sum_{\ell}\mathbf{G}_{\ell}(\mathbf{q}),\text{ where }\mathbf{G}_{\ell}(\mathbf{q})=\mathbf{J}_{\ell}^{T}(\mathbf{q})\left\{ \mathbb{V}_{\mathbf{\pi}_{\ell}}[\mathbf{x}_{\ell}|\boldsymbol{\theta}_{\ell}](\boldsymbol{\psi}(\mathbf{q}))\right\} \mathbf{J}_{\ell}(\mathbf{q}),\label{eq:the_metric_tensor}
\end{equation}
as the metric tensor. Note that $\mathbf{G}(\mathbf{q})$ is the sum
of induced (pull-back) pseudo-metric tensors from (the domain of)
$(\mathbf{x}_{\ell}|\boldsymbol{\theta}_{\ell})$ to the sampling
space characterized by $\mathbf{q}$. 

Before proceeding, some remarks are in order. 
\begin{itemize}
\item Under certain additional assumptions, (certain rows/columns of) $\mathbf{G}_{\ell}$
may itself be interpreted as a LGC. More precisely, momentarily assuming
that there exist subsets $I$ and $J$ of $\{1,\dots,d\}$ where $I\cap J=\emptyset$,
functions $\mathbf{a},\mathbf{c}$ and invertible matrix $\mathbf{B}$
so that $\mathbf{\boldsymbol{\phi}}_{\ell}(\mathbf{q})=\mathbf{a}(\mathbf{q}_{J})+\mathbf{B}\mathbf{q}_{I}$
and $\boldsymbol{\psi}_{\ell}(\mathbf{q})=\mathbf{c}(\mathbf{q}_{J})$.
Then it follows from (\ref{eq:repar_LCG}, with $\boldsymbol{\Psi}=\mathbf{c}$)
that the $I\cup J$ -rows/columns of $\mathbf{G}_{\ell}$ is the LGC
of the density of $\mathbf{q}_{I}|\mathbf{q}_{J}$ implied by $\pi_{\ell}$,
$\boldsymbol{\phi}_{\ell}$ and $\boldsymbol{\psi}_{\ell}$.
\item Informally speaking, (\ref{eq:the_metric_tensor}) correctly represents
the precision matrix of any (possibly degenerate) linear Gaussian
structure. More precisely, in the case characterized by $\pi_{\ell}=\mathcal{N}(\mathbf{a}+\mathbf{A}\mathbf{q}_{I}|\mathbf{b}+\mathbf{B}\mathbf{q}_{I},\mathbf{P}^{-1}(\mathbf{q}_{J}))$,
where again $I\cap J=\emptyset$, and $\mathbf{A},\mathbf{B}$ not
necessarily invertible, the $I$-rows/columns of $\mathbf{G}_{\ell}$
are equal to $-\nabla_{\mathbf{q}_{I}}^{2}\log\pi_{\ell}$. 
\item In general, (\ref{eq:target_rep}) admit non-linear $\boldsymbol{\phi}_{\ell}$s,
and $\mathbf{G}_{\ell}$ will be a SPSD matrix also in the case of
such non-linear $\boldsymbol{\phi}_{\ell}$s. Inclusion of this possibility
is mainly done as the automatic implementation of (\ref{eq:the_metric_tensor})
(see Section \ref{subsec:Computing_G}) does not distinguish between
linear and non-linear $\boldsymbol{\phi}_{\ell}$s. Still, the interpretation
of non-linear $\boldsymbol{\phi}_{\ell}$s is not obvious as (\ref{eq:target_rep})
does not involve the log-Jacobian determinants of the $\boldsymbol{\phi}_{\ell}$s.
Hence, the use of non-linear $\boldsymbol{\phi}_{\ell}$s is advised
against. Further, if $\boldsymbol{\phi}_{\ell}$ is bijective from
some subset of $\mathbf{q}$, the need for non-linear $\boldsymbol{\phi}_{\ell}$s
may be alleviated by choosing an equivalent base distribution on the
pre-image of $\boldsymbol{\phi}_{\ell}$.
\item For models not involving latent variables so that $\mathbf{q}$ may
be interpreted as a parameter vector, (\ref{eq:the_metric_tensor})
reduces to (\ref{eq:GC_MT}) with the modification the matrix $\mathcal{R}$
is now the sum of prior argument log-gradient covariances $\mathcal{V}_{\ell}$
for $\ell$s corresponding to prior terms in (\ref{eq:target_rep}).
Further, for terms corresponding to the log-likelihood function, say
$\ell\in\mathcal{L}$, $\sum_{\ell\in\mathcal{L}}\mathbf{G}_{\ell}(\mathbf{q})$
is exactly the Fisher of the observations with respect to the parameter
$\mathbf{q}$. In this sense, the proposed methodology may be seen
as a generalization of the suggestion of \citet{girolami_calderhead_11}
to a much more general class of models involving latent variables.
\item For a given model with target distribution $\pi(\mathbf{q})$, the
metric tensor (\ref{eq:the_metric_tensor}) is in general \emph{not}
invariant to how the factorization (\ref{eq:target_rep}) is carried
out. As an example, consider the simple funnel-model $q_{1}\sim N(0,1),\;q_{2}|q_{1}\sim N(0,\exp(-3q_{1}))$.
The factorization (\ref{eq:target_rep}) may be done as $\pi(\mathbf{q})=\pi_{1}(q_{1})\pi_{2}(q_{2}|q_{1})$
which leads to $\mathbf{G}(\mathbf{q})=\text{diag}(11/2,\exp(-3q_{1}))$,
or simply $\pi(\mathbf{q})=\pi_{1}(q_{1},q_{2})$ which lead to $\mathbf{G}(\mathbf{q})=\mathbb{V}_{\pi_{1}}[q_{1},q_{2}|]=\text{diag}(11/2,\exp(-9/2))$.
Clearly, the former metric tensor provides useful scaling information
for sampling from the funnel distribution, whereas the latter metric
tensor is not useful in this case. From this simple example, it seems
advisable if possible, to factorize highly non-Gaussian joint distributions
rather than to derive and use the LGC of the joint distribution. In
fact, working with a single factor with $\phi_{1}=\mathbf{q}$ in
(\ref{eq:target_rep}) would, as exemplified by the latter metric
tensor above, result in a constant (Euclidean) metric tensor. The
factorization issue is further explored in Section \ref{sec:A-Wishart-transition_SV}.
\end{itemize}

\subsection{Examples}

To get a sense of the workings of the proposed methodology, some small
examples are considered here. 
\begin{figure}

\centering{}\includegraphics[scale=0.45]{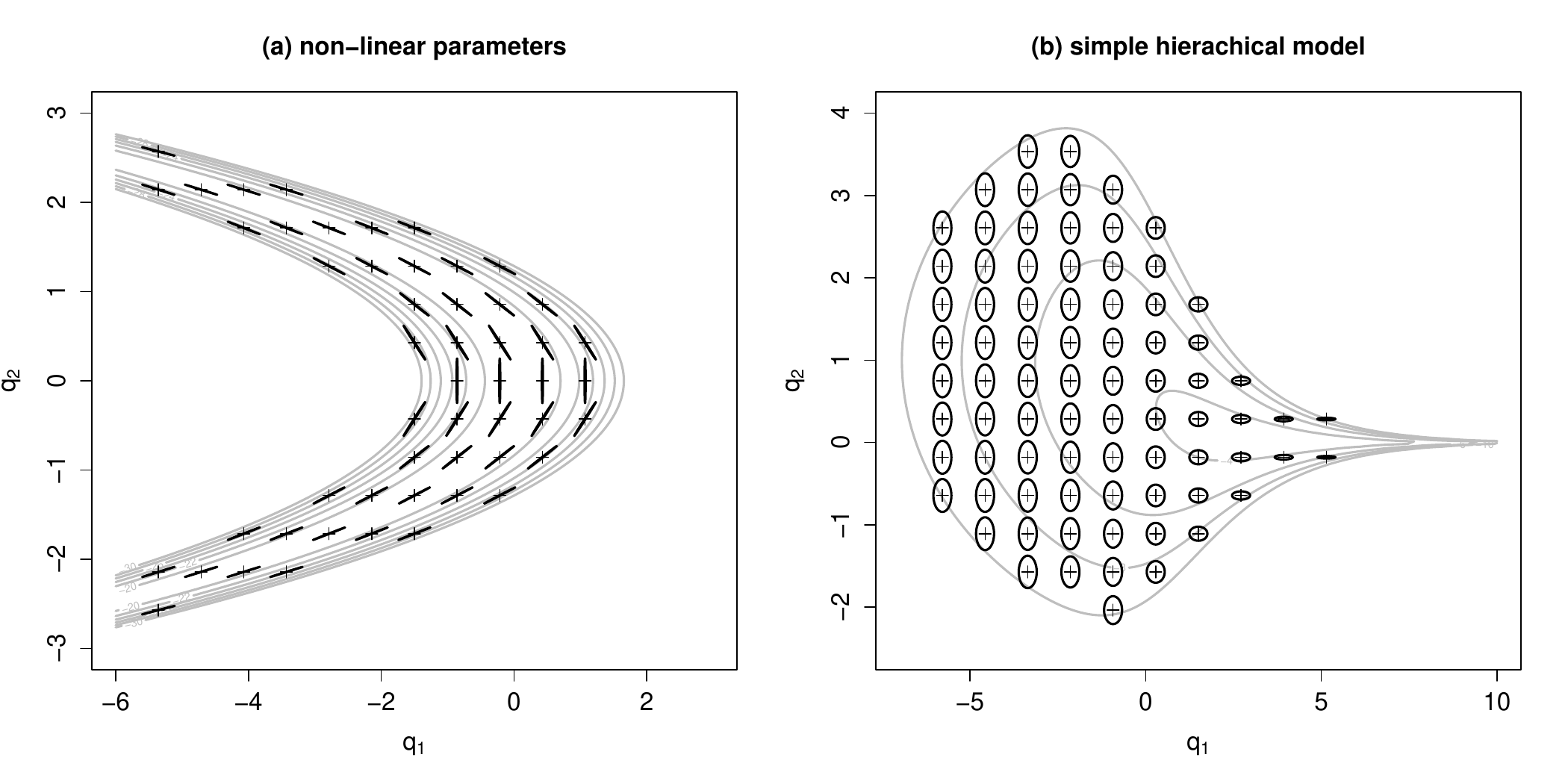}\caption{\label{fig:toy_models}Illustrations of the proposed metric tensor
for (a) : model (\ref{eq:BC_model}) and (b) : model (\ref{eq:obs_funnel})
with $y=1$. In both plots, gray lines indicated target log-density
contours, whereas the black closed curves are equi-probability ellipsis'
associated with $N(\mathbf{q},\mathbf{G}^{-1}(\mathbf{q}))$ for values
of $\mathbf{\mathbf{q}}$ indicated by +-signs. }
\end{figure}

\subsubsection{Non-linear parameter transformations}

First, a non-hierarchical model similar to that of \citet{born_cornebise_11}
is considered. The model involves two parameters $\mathbf{q}=(\theta_{1},\theta_{2})$
and may be summarized by 
\begin{equation}
\ell=1,\dots,n\;:\;y_{\ell}|\theta_{1},\theta_{2}\sim N(\theta_{1}+\theta_{2}^{2},1)\text{ and }\ell=n+1\;:\;(\theta_{1},\theta_{2})\sim N(\mathbf{0},100\mathbf{I}_{2}).\label{eq:BC_model}
\end{equation}
Then (\ref{eq:the_metric_tensor}) result in 
\begin{equation}
\mathbf{G_{\ell}}=\left[\begin{array}{cc}
1 & 2\theta_{2}\\
2\theta_{2} & 4\theta_{2}^{2}
\end{array}\right],\ell=1,\dots,n,\;\text{and }\mathbf{G}_{n+1}=100^{-1}\mathbf{I}_{2}.\label{eq:tensor_BC}
\end{equation}
The resulting metric tensor is the same as (\ref{eq:GC_MT}) obtained
in \citet{born_cornebise_11}, which owes to the fact that for the
Gaussian prior $\pi_{n+1}$, the $\mathcal{V}_{n+1}$ is equal to
the negative Hessian at the mode $\mathcal{R}$. In Figure \ref{fig:toy_models}
(a), it is seen that (\ref{eq:tensor_BC}) appears to accurately represent
the local scaling properties of the target distributions.

\subsubsection{A simple hierarchical model\label{subsec:A-simple-hierarchical}}

Now consider a simple hierarchical model with $\mathbf{q}=(\lambda,z)$
where $\lambda$ is the a-priori log-precision of the latent variable
$z$. The model is characterized by 
\begin{equation}
\ell=1\;:\;\lambda\sim N(0,3^{2}),\;\ell=2\;:\;z|\lambda\sim N(0,\exp(-\lambda))\text{ and }\ell=3\;:\;y|z\sim N(z,1).\label{eq:obs_funnel}
\end{equation}
In this case, (\ref{eq:the_metric_tensor}) results in
\begin{equation}
\mathbf{G}_{1}=\left[\begin{array}{cc}
3^{-2} & 0\\
0 & 0
\end{array}\right],\;\mathbf{G}_{2}=\left[\begin{array}{cc}
\frac{1}{2} & 0\\
0 & \exp(\lambda)
\end{array}\right],\;\mathbf{G}_{3}=\left[\begin{array}{cc}
0 & 0\\
0 & 1
\end{array}\right].\label{eq:tensor_hierar}
\end{equation}
Figure \ref{fig:toy_models} (b) illustrates (\ref{eq:tensor_hierar}),
where it is seen that the scaling properties of the target distribution
appears well represented.

\subsubsection{Intrinsic Gaussian}

In the final small example, consider the intrinsic Gaussian model
\citep[see e.g. ][]{rue_held_05} for $\mathbf{q}\in\mathbb{R}^{3}$
characterized by 
\begin{equation}
\ell=1\;:\;q_{1}-q_{2}\sim N(0,\kappa^{-1}),\;q_{1}-q_{3}\sim N(0,\kappa^{-1})\text{ and }q_{2}-q_{3}\sim N(0,\kappa^{-1}),\label{eq:toy_gmrf}
\end{equation}
for some fixed $\kappa$. This formulation still fits directly into
(\ref{eq:target_rep}), even with non-invertible argument functions,
e.g. $\phi_{1}(\mathbf{q})=q_{1}-q_{2}$ and with $\pi_{1}$ corresponding
to $\mathcal{N}(x|0,\kappa^{-1})$. Further, in line with the second
comment of Section \ref{subsec:Proposed-metric-tensor}, the metric
tensor (\ref{eq:the_metric_tensor}), 
\[
\mathbf{G}=\left[\begin{array}{ccc}
2\kappa & -\kappa & -\kappa\\
-\kappa & 2\kappa & -\kappa\\
-\kappa & -\kappa & 2\kappa
\end{array}\right],
\]
is the (degenerate) precision matrix associated with (\ref{eq:toy_gmrf}).

\subsection{Automatic computation of $\mathbf{G}(\mathbf{q})$\label{subsec:Computing_G}}

\begin{table}
\begin{centering}
\begin{tabular}{clcl}
\hline 
Line \# & Code &  & Comments\tabularnewline
\hline 
1 & \texttt{\footnotesize{}PARAMETER\_SCALAR(lambda);} &  & Stores $\lambda$ as an \texttt{\footnotesize{}amtVar} variable and
sets $\nabla_{\mathbf{q}}\lambda=[1,0]$.\tabularnewline
2 & \texttt{\footnotesize{}PARAMETER\_SCALAR(z);} &  & Stores $z$ as an \texttt{\footnotesize{}amtVar} variable and sets
$\nabla_{\mathbf{q}}z=[0,1]$.\tabularnewline
3 & \texttt{\footnotesize{}model\_\_ += normal\_ld(lambda,0.0,3.0);} &  & Computes $\log\pi_{1}=\log\mathcal{N}(\lambda|0,3^{2})$ and $\mathbf{G}_{1}=[\nabla_{\mathbf{q}}\lambda]^{T}3^{-2}\left[\nabla_{\mathbf{q}}\lambda\right]$.\tabularnewline
4 & \texttt{\footnotesize{}amtVar sigma = exp(-0.5{*}lambda);} &  & Computes $\sigma=\exp(-\frac{\lambda}{2})$ and sets $\nabla_{\mathbf{q}}\sigma=-\frac{1}{2}\exp(-\frac{\lambda}{2})\nabla_{\mathbf{q}}\lambda$.\tabularnewline
\multirow{2}{*}{5} & \multirow{2}{*}{\texttt{\footnotesize{}model\_\_ += normal\_ld(z,0.0,sigma);}} &  & Computes $\log\pi_{2}=\log\mathcal{N}(z|0,\sigma^{2})$, $\mathbf{J}_{2}=[[\nabla_{\mathbf{q}}z]^{T},[\nabla_{\mathbf{q}}\sigma]^{T}]^{T}$\tabularnewline
 &  &  & and $\mathbf{G}_{2}=\mathbf{J}_{2}^{T}\text{diag}(\sigma^{-2},2\sigma^{-2})\mathbf{J}_{2}$.\tabularnewline
6 & \texttt{\footnotesize{}model\_\_ += normal\_ld(1.0,z,1.0);} &  & Computes $\log\pi_{3}=\log\mathcal{N}(1|z,1)$ and $\mathbf{G}_{3}=[\nabla_{\mathbf{q}}z]^{T}[1][\nabla_{\mathbf{q}}z]$.\tabularnewline
\hline 
\end{tabular}
\par\end{centering}
\caption{\label{tab:An-implementation-of}An implementation of the model in
section \ref{subsec:A-simple-hierarchical} using the \texttt{amt}
library, were $\mathbf{G}_{1},\;\mathbf{G}_{2}$ and $\mathbf{G}_{3}$
are given in (\ref{eq:tensor_hierar}) and $\mathbf{q}=(\lambda,z)^{T}$.}
\end{table}
Equation \ref{eq:the_metric_tensor} may at first glance appear somewhat
intimidating to compute for a general non-linear model (\ref{eq:target_rep}).
However, it may be computed in a completely automatic fashion based
on AD for distribution families $\{\pi_{\ell}(\mathbf{x}|\boldsymbol{\theta})\}_{\ell}$
all having known LGCs \{$\mathbb{V}_{\pi_{\ell}}\}_{\ell}$ (or Fisher
information matrices in cases of discretely distributed observations).
More specifically, the proposed methodology leading to (\ref{eq:the_metric_tensor})
has been implemented in the C++ library \texttt{amt}, which is a part
of the pdmphmc package (\url{https://github.com/torekleppe/pdmphmc})
which will be described shortly. With access to library \texttt{amt},
the user is only responsible for providing C++ code for specifying
the model (\ref{eq:target_rep}). 

As an illustration, a working implementation of the model in section
\ref{subsec:A-simple-hierarchical} using library \texttt{amt} is
given in the code column of Table \ref{tab:An-implementation-of}.
The computations arriving at (\ref{eq:tensor_hierar}) are done in
an automatic manner as described in the comments column of Table \ref{tab:An-implementation-of}.
The library consist firstly of an AD type \texttt{amtVar}, which is
used to store both the value and also the gradient (w.r.t. $\mathbf{q}$)
of every quantity in the model that depends on $\mathbf{q}$. The
\texttt{amtVar} type is based on an implementation of first order
forward mode AD, which is sparse in the sense that it only stores
the non-zero elements of each given gradient. This practice (as opposed
to the celebrated backward mode AD) is informed by the fact that the
rows of $\mathbf{J}_{\ell}$ are typically very sparse (each parameter/latent
variable plays only a limited number of ``roles'') for hierarchical
models. In Table \ref{tab:An-implementation-of}, lines 1,2 and 4
illustrates the how the \texttt{amtVar} type is used to maintain the
gradient of $\lambda,\;z$ and $\sigma=\exp(-\lambda/2)$ with respect
to $\mathbf{q}$. 

Secondly, the \texttt{amt} library consist of a collection of probability
distributions with known LGCs, illustrated here by the univariate
Gaussian distribution \texttt{normal\_ld()} in Table \ref{tab:An-implementation-of},
lines 3, 5 and 6. Whenever such a function is called, the posterior
log density kernel (\ref{eq:target_rep}) is incremented by the appropriate
$\log\pi_{\ell}$. Further, if the function is called with arguments
and/or parameters of \texttt{amtVar} type, (\ref{eq:the_metric_tensor})
is incremented by the appropriate $\mathbf{G}_{\ell}$, computed from
the LGC of the distribution and the gradients of the arguments and/or
parameters.

The methodology for computing log-densities $\left\{ \log\pi_{\ell}\right\} _{\ell}$,
LGCs $\{\mathbb{V}_{\pi_{\ell}}\}_{\ell}$, Jacobians $\{\mathbf{J}_{\ell}\}_{\ell}$
and the Cholesky-factorization of the resulting $\mathbf{G}(\mathbf{q})$
(based on either dense- or sparse \citep{davis_sparse} storage) required
to compute (\ref{eq:hamiltonian}) are in turn differentiated using
the backward mode Stan AD \citep{JSSv076i01} to obtain the gradient
in (\ref{eq:ham_eq_2}). 

From a computational performance perspective, it is still advisable
to keep the parameter functions $\{\mathbf{\boldsymbol{\psi}}_{\ell}(\mathbf{q})\}_{\ell}$
as simple as possible to avoid lengthy forward mode AD Jacobian calculations.
This may be accomplished by defining new base distributions based
on (\ref{eq:repar},\ref{eq:repar_LCG}, with $\mathbf{a}=\mathbf{0}$,
$\mathbf{B}=\mathbf{I}$) taking into account the re-parameterization
represented by $\boldsymbol{\Psi=}\boldsymbol{\psi}_{\ell}$. As an
example, consider a model involving a normal linear model $\mathbf{y}\sim N(\mathbf{X}\boldsymbol{\beta},\sigma^{2}\mathbf{I})$
with constant design matrix $\mathbf{X}$ and $\boldsymbol{\beta}=\mathbf{q}_{I}$
for some index set $I$. Then it would be more effective to use the
base distribution $\mathbf{y}|(\boldsymbol{\beta},\sigma)$ with $\boldsymbol{\psi}(\mathbf{q})=(\mathbf{q}_{I},\sigma(\mathbf{q}))$,
rather than base distribution $\mathbf{y}|(\boldsymbol{\mu},\sigma)$
with $\boldsymbol{\psi}(\mathbf{q})=(\mathbf{X}\mathbf{q}_{I},\sigma(\mathbf{q}))$.
This follows from that large savings may be realized by pre-computing
the factor $\mathbf{X}^{T}\mathbf{X}$ of the LGC of $\mathbf{y}|(\boldsymbol{\beta},\sigma)$.
In the $\mathbf{y}|(\boldsymbol{\mu},\sigma)$, $\boldsymbol{\psi}(\mathbf{q})=(\mathbf{X}\mathbf{q}_{I},\sigma(\mathbf{q}))$
case, on the other hand, both calculating the non-trivial Jacobian
of $\psi$ using the forward mode AD routines, and also calculating
the matrix product (\ref{eq:the_metric_tensor}) would have to be
repeated for each evaluation of $\mathbf{G}$. Deriving and implementing
LGCs for the most common non-trivial ``submodels'' is an ongoing
effort.

\section{Examples\label{sec:Examples}}

This section considers real data example problems chosen in order
to illustrate several aspects of the proposed methodology. In addition
to the examples below, a further example, a mixed effects model for
the Salamander mating data of \citet[Chapter 14.5]{mccu:neld:1989}
may be found in Appendix \ref{sec:Salamander-mating-model}. For this
model, due to rather high CPU time usage, the proposed methodology
does not outperform the benchmark Euclidean metric sampler.

\subsection{Implementation details}

The examples are all implemented using the pdmphmc C++ library (development
version available at \url{https://github.com/torekleppe/pdmphmc})
which provides both Riemann manifold (RM)- and Euclidean metric (EM)
numerical GRHMC (NGRHMC) processes, along with the library for automatic
computation of (\ref{eq:the_metric_tensor}) and an interface to R
to facilitate building and running of models. 

A snapshot of the version of pdmphmc used in this paper, along with
R code, data sets etc used in this paper is also available at \url{https://github.com/torekleppe/AMTpaperCode}.

If not otherwise mentioned, in all the cases below, RM NGRHMC processes,
along with EM NGRHMC processes as reference, are run with $T_{\max}=10,000$
with the former half discarded as burn in. The ``sampling''-part
of the trajectories are sampled at 1000 equidistant times. In all
cases, 8 independent trajectories were used, and the reported figures
are calculated across these 8 trajectories. The 8 trajectories were
run in parallel in 2 batches of 4 trajectories on a 2020 macbook pro.
The reported CPU times are the sum across trajectories of the CPU
times required for generating the ``sampling''-parts of the trajectories.
The effective sample sizes (ESS) and (modified) Gelman-Rubin $\hat{R}$-statistics
\citep{GelmanBDA3} are calculated using the \texttt{rstan::monitor()}-function
\citep{Rstan_ref}. Note that the reported results involve more trajectories
than one would use in a typical application of the methodology, in
order to reliably compare ESSes and time-weighted ESSes across sampling
methods. E.g., for a typical application one would rather use say
4 trajectories computed in parallel (resulting in wall-clock time
being $1/8$ of reported CPUtimes) to obtain roughly half the reported
ESSes.

As a further benchmark, the examples were also implemented and sampled
using Stan through the R interface rstan (version 2.26.23 with StanHeaders
version 2.26.28) on the same computer. In all cases, 8 chains of 1000
transitions (post warmup) and otherwise default settings were used
for Stan.

\subsection{Zero-inflated Poisson mixed regression}

The first model considered is a mixed effect regression model with
zero-inflated Poisson count responses. Specifically, for a response
$y_{i}$ the response distribution is given by
\begin{align}
P(y_{i} & =0|\eta_{i},g_{i})=\frac{\exp(g_{i})+\exp(-\exp(\eta_{i}))}{1+\exp(g_{i})},\label{eq:ZIPoisson_1}\\
P(y_{i} & =x|\eta_{i},g_{i})=\frac{\exp(x\eta_{i}-\exp(\eta_{i}))}{(1+\exp(g_{i}))x!},\;x=1,2,\dots\label{eq:ZIPoisson_2}
\end{align}
which may be interpreted as mixture of a point-mass in $y_{i}=0$
and a Poisson distribution with mean $\exp(\eta_{i})$, were the mixture
weight of the $y_{i}=0$ point mass is $\exp(g_{i})(1+\exp(g_{i}))^{-1}$.
Consequently, $E(y_{i}|\eta_{i},g_{i})=\exp(\eta_{i})(1+\exp(g_{i}))^{-1}.$
The Fisher information of (\ref{eq:ZIPoisson_1},\ref{eq:ZIPoisson_2})
with respect to $(\eta_{i},g_{i})$ has closed (but complicated) form
and is given in Appendix \ref{subsec:Details-for-Poisson}.
\begin{table}

\centering{}%
\begin{tabular}{cccccccccc}
\hline 
 & CPU time & $\max\hat{R}$ & \multicolumn{3}{c}{$\sigma$} &  & $\boldsymbol{\beta}_{\eta}$ & $\boldsymbol{\beta}_{g}$ & $\mathbf{b}$\tabularnewline
\cline{4-6} \cline{5-6} \cline{6-6} 
 & (s) &  & post. & post. & ESS &  & min & min & min\tabularnewline
 &  &  & mean & SD &  &  & ESS & ESS & ESS\tabularnewline
\hline 
\multirow{2}{*}{RM} & 7178 & 1.002 & 1.37 & 0.21 & 5811 &  & 5894 & 10537 & 5557\tabularnewline
 &  &  &  &  & {[}0.8{]} &  & {[}0.8{]} & {[}1.5{]} & {[}0.8{]}\tabularnewline
\hline 
\multirow{2}{*}{EM} & 1034 & 1.002 & 1.37 & 0.22 & 7740 &  & 4367 & 514 & 2545\tabularnewline
 &  &  &  &  & {[}7.5{]} &  & {[}4.2{]} & {[}0.5{]} & {[}2.5{]}\tabularnewline
\hline 
\multirow{2}{*}{Stan} & 65 & 1.008 & 1.38 & 0.22 & 5522 &  & 1095 & 985 & 1166\tabularnewline
 &  &  &  &  & {[}85{]} &  & {[}17{]} & {[}15{]} & {[}17{]}\tabularnewline
\hline 
\end{tabular}\caption{\label{tab:Results-for-the-salamander-count}Results for the zero-inflated
Poisson mixed regression for Salamander count data. CPU time is the
total computing time spent sampling (post warmup) by 8 independent
trajectories. $\max\hat{R}$ is the maximum $\hat{R}$-statistic of
over all quantities that are sampled. The table provides the posterior
mean and standard deviation, and ESS of the random effects variance
parameter $\sigma$, in addition to the worst-case ESS and CPU time
weighted ESS (measured ESS per second CPU time) in {[}{]}-brackets.}
\end{table}

The data set considered is the Salamander data set originally discussed
by \citet{https://doi.org/10.1111/1365-2664.12585} which is included
in the R-package \texttt{glmmTMB} \citep[see][]{RJ-2017-066} and
consist of $n=644$ observations. The ``mean'' linear predictor
$\boldsymbol{\eta}$ involves a total of 7 fixed effects (including
an intercept term, with corresponding parameter $\boldsymbol{\beta}_{\eta}\in\mathbb{R}^{7}$)
and a total of 23 random effects $\mathbf{b}\in\mathbb{R}^{23}$ with
common variance parameter $\sigma^{2}$. The ``zero inflation''
linear predictor $\mathbf{g}$ consist of the same 7 fixed effects
(with corresponding parameter $\boldsymbol{\beta}_{g}\in\mathbb{R}^{7}$).
Due to a complicated sparsity structure in $\mathbf{G}(\mathbf{q})$,
and the moderate dimension of $\mathbf{q}=(\log(\sigma^{2}),\mathbf{b},\boldsymbol{\beta}_{\eta},\boldsymbol{\beta}_{g})$,
dense storage of $\mathbf{G}(\mathbf{q})$ was used as it resulted
in slightly better performance. Further details are provided in Appendix
\ref{subsec:Details-for-Poisson}. 

Table \ref{tab:Results-for-the-salamander-count} provides results
for the RM, EM and Stan samplers. The $\hat{R}$-statistics indicated
that all samplers exhibits satisfactory mixing. It is seen that RM
sampler is substantially slower than EM in terms of simulating the
same (process time) amount of trajectory, owing to that each evaluation
of Hamilton\textasciiacute s equations is substantially more costly
than for the EM counterpart. Further, Stan uses an order of magnitude
less time than EM. The ESS of $\sigma$ along with the worst case
ESSes across $\boldsymbol{\beta}_{\eta}$, $\boldsymbol{\beta}_{g}$
and $\mathbf{b}$ are similar for RM and EM except for a much poorer
ESS for $\boldsymbol{\beta}_{g}$ for the EM sampler. This shortfall
is likely to be related to the non-linear interaction between $\boldsymbol{\beta}_{g}$
and the remaining sampled quantities. The RM sampler, on the other
hand, exhibit no such inefficiencies indicating that the proposed
metric tensor is able to reflect these interactions. Even if Stan
has relatively moderate raw ESSes for $\boldsymbol{\beta}_{\eta},\boldsymbol{\beta}_{g}$
and $\mathbf{b}$, the very small CPU time of Stan result in the smallest
time-weighted ESS for all parameters/latent variables for this model.

\subsection{Random walk stochastic volatility with leverage effect}

\begin{table}
\centering{}%
\begin{tabular}{ccccccc}
\hline 
 & CPU time & $\max\hat{R}$ & $\rho$ & $\sigma$ & $z_{0}$ & $z_{T}$\tabularnewline
 & (s) &  & ESS & ESS & ESS & ESS\tabularnewline
\hline 
\multirow{2}{*}{RM} & 9397 & 1.006 & 1762 & 1864 & 11240 & 13306\tabularnewline
 &  &  & {[}0.19{]} & {[}0.20{]} & {[}1.20{]} & {[}1.42{]}\tabularnewline
\hline 
\multirow{2}{*}{EM} & 9238 & 1.101 & 60 & 1767 & 8809 & 4143\tabularnewline
 &  &  & {[}0.01{]} & {[}0.19{]} & {[}0.95{]} & {[}0.45{]}\tabularnewline
\hline 
\end{tabular}\caption{\label{tab:Effective-sample-sizes_SV-lev}Effective sample sizes and
diagnostics for the stochastic volatility model with leverage effect
(\ref{eq:SV_transition},\ref{eq:SV_obs}) applied to a data set of
S\&P500 log-returns. }
\end{table}
Next, a random walk stochastic volatility (SV) model with leverage
effect \citep[see e.g. ][]{yu05} is considered. The latent log-volatility
$\mathbf{z}=(z_{0},\dots,z_{T})$ evolves according to a Gaussian
random walk
\begin{equation}
z_{t}|z_{t-1},\sigma\sim N(z_{t-1},\sigma^{2}),\;t=1,\dots,T.\label{eq:SV_transition}
\end{equation}
Further, the log-return observations $\mathbf{y}=(y_{1},\dots,y_{T})$
are modeled as
\begin{equation}
y_{t}|z_{t},z_{t-1},\rho,\sigma\sim N\left(\rho\exp\left[\frac{z_{t-1}}{2}\right]\frac{z_{t}-z_{t-1}}{\sigma},\exp(z_{t-1})(1-\rho^{2})\right),\;t=1,\dots,T.\label{eq:SV_obs}
\end{equation}
Finally, the priors $\rho\sim$ Uniform$(-1,1)$ and $\sigma^{2}\sim0.1/\chi_{10}^{2}$
completes the model. The data set consisted of $T=2515$ log-return$\times100$
observations of the S\&P500 index spanning Oct. 1st 1999 to Sep. 30th
2009 \citep[previously used by][]{1601.01125}. Note that scale/covariance
of conditional posterior $\mathbf{z}|\mathbf{y},\rho,\sigma$ depends
non-linearly on both $\rho$ and $\sigma$, i.e. the posterior will
be ``funnel shaped along two separate dimensions''. Consequently,
posterior sampling may be troublesome for many MCMC methods, and the
model may be considered a rather challenging one.

For the RM variant of the sampler, the LGCs of $\mathbf{z}|\sigma$
consistent with (\ref{eq:SV_transition}) and $y_{t}|z_{t},z_{t-1},\rho,\sigma,\;t=1,\dots,T$
consistent with (\ref{eq:SV_obs}) were used. In addition, otherwise
identical calculations were done based on the LGC $\mathbb{V}[x|(\mu,\sigma)]$
for the $x\sim N(\mu,\sigma^{2})$-distribution for the observation
equation (\ref{eq:SV_obs}) (and hence parameter functions $\boldsymbol{\psi}=(\mu,\sigma)=(\rho\exp\left[\frac{z_{t-1}}{2}\right]\frac{z_{t}-z_{t-1}}{\sigma},\sqrt{\exp(z_{t-1})(1-\rho^{2})})$).
The latter approach, where non-trivial calculations are done using
the general-purpose forward mode AD system, leads to an increase in
computing time by around 30\% (but otherwise identical results, hence
not reported). Sparse storage with variable ordering $\mathbf{q}=(\mathbf{z},-1+2\text{logit}(\rho),\log(\sigma))$
was used, so that the sparsity structure has an arrowhead shape, which
lends itself well to the spare Cholesky factorization. For the EM
and Stan implementations, the sampled quantity corresponding to the
latent variable was $\bar{\mathbf{z}}_{0:T}=(z_{0},(z_{1}-z_{0})\sigma^{-1},\dots,(z_{T}-z_{T-1})\sigma^{-1})$
(i.e. so that $\bar{\mathbf{z}}_{1:T}\sim N(\mathbf{0},\mathbf{I}_{T})$
a priori) rather than $\mathbf{z}$ in order to reduce ``funnel''
effects determined by $\sigma$.

Table \ref{tab:Effective-sample-sizes_SV-lev} provides effective
sample sizes and other diagnostic information for the EM and RM samplers,
whereas Stan failed to produce meaningful results, issued a large
number of warning messages, and is hence not reported on. From Table
\ref{tab:Effective-sample-sizes_SV-lev}, it is seen that computing
times are roughly equal, whereas the EM sampler fails to properly
explore the posterior distribution of $\rho$. This failure is likely
to be related to that the $\bar{\mathbf{z}}$-parameterization does
not take into account how the scale of $\bar{\mathbf{z}}|\mathbf{y},\sigma,\rho$
varies with $\rho$. No such deficiencies are seen for the RM sampler,
as the dependence of the scale of $\mathbf{z}|\mathbf{y},\sigma,\rho$
on the parameters $(\sigma,\rho)$ is automatically accounted for
in the metric tensor.

\subsection{CEV model with additive noise}

\begin{table}
\centering{}%
\begin{tabular}{ccccc}
\hline 
 & Post. & Post. & \multicolumn{2}{c}{ESS}\tabularnewline
 & mean & SD &  & \tabularnewline
\hline 
$\alpha$ & 0.010 & 0.009 & 6163 & {[}0.31{]}\tabularnewline
$\beta$ & 0.171 & 0.174 & 6186 & {[}0.31{]}\tabularnewline
$\sigma_{x}$ & 0.404 & 0.061 & 7695 & {[}0.39{]}\tabularnewline
$\gamma$ & 1.180 & 0.060 & 6914 & {[}0.35{]}\tabularnewline
$\sigma_{y}$ & 0.00054 & $2.3\times10^{-5}$ & 3698 & {[}0.19{]}\tabularnewline
$x_{1}$ & 0.095 & 0.0005 & 6419 & {[}0.32{]}\tabularnewline
$x_{T}$ & 0.061 & 0.0005 & 6273 & {[}0.31{]}\tabularnewline
\hline 
\end{tabular}\caption{\label{tab:Posterior-distributions-and-CEV}Posterior distributions
and diagnostics information CEV model with additive noise model (\ref{eq:CEV_latent}-\ref{eq:CEV_obs}).
Total CPU time for the 8 trajectories was 19933 seconds, and max $\hat{R}$
for the stored states (parameters and $x_{1}$,$x_{T}$) was 1.001977. }
\end{table}
This section considers a daily time-discretization of constant elasticity
of volatility model \citep{chan_et_al_1992} with additive Gaussian
noise for interest rate data previously considered by \citet{Kleppe2017}.
The model is formulated in continuous time with unit of continuous
time being one year, and time-discretized to (business day) daily
observations with time steps $\Delta=1/252$. The model may be summarized
by the time-discretized non-linear latent ``true'' short term interest
rate

\begin{align}
x_{t} & =x_{t-1}+\Delta(\alpha-\beta x_{t-1})+\sigma_{x}\sqrt{\Delta}x_{t-1}^{\gamma}\varepsilon_{t},\;\varepsilon_{t}\sim\text{ iid }N(0,1),\;t=2,\dots,T,\label{eq:CEV_latent}\\
x_{1} & \sim N(0.09569,0.01^{2}),\label{eq:CEV_first_latent}
\end{align}
and the daily observations contaminated with additive Gaussian noise:
\begin{equation}
y_{t}=x_{t}+\sigma_{y}\eta_{t},\;\eta_{t}\sim\text{ iid }N(0,1),\;t=1,\dots,T.\label{eq:CEV_obs}
\end{equation}
The data set considered was $T=3082$ observations of the 7-day Eurodollar
deposit spot rates from January 2, 1983, to February 25, 1995 previously
used by \citet{aitsahalia1996,Kleppe2017}. Further details, including
priors may be found in Appendix \ref{subsec:Details-related-to-CEV},
and were chosen to be identical to the setup of \citet{Kleppe2017}
to allow for comparison with modified Cholesky Riemann manifold HMC.

The proposed methodology was implemented using univariate Gaussian
LGCs with the standard parameterization, i.e. $\mathbb{V}[x|(\mu,\sigma)]$
for the $x\sim N(\mu,\sigma^{2})$-distribution, (and not a bespoke
LGC for say $x_{t}|x_{t-1},\alpha,\beta,\sigma_{x},\gamma$ in the
case of (\ref{eq:CEV_latent})) thus relying on the general purpose
forward mode AD system to handle the non-linear relations between
the sampled quantities. Using the variable ordering $\mathbf{q}=(\mathbf{z},\alpha,\beta,\log(\sigma_{x}^{2}),\gamma,\log(\sigma_{y}^{2}))$,
the metric tensor $\mathbf{G}(\mathbf{q})$ again has an arrow head
sparsity structure which lend itself well to the sparse Cholesky factorization
used.

Only results for RM based sampler are presented in Table \ref{tab:Posterior-distributions-and-CEV},
as direct EM-based or Stan-based sampling methods for this model failed
to be even remotely competitive/produce reliable results, and dynamic
rescaling methods for EM/Stan are not directly applicable due to the
non-linear nature of (\ref{eq:CEV_latent}). Table \ref{tab:Posterior-distributions-and-CEV}
indicate that the proposed methodology produces reliable output with
ESSes being quite even across the reported dimensions. 

As a benchmark for sampling efficiency, \citet{Kleppe2017} reports
sampling efficiencies about an order of magnitude slower than those
reported in between Table \ref{tab:Posterior-distributions-and-CEV}
(0.062 and 0.035 ESS per second) for a Riemann manifold HMC method
(based on reversible symplectic integrator). Disentangling the effect
of the here proposed metric tensor versus the modified Cholesky applied
to Hessian approach of \citet{Kleppe2017}, from the effect of different
ODE integration strategies is impossible based on this information.
Still, the combination of Riemann manifold NGRHMC processes and the
here proposed metric tensor is highly competitive while at the same
time requiring minimal expertise and coding efforts from the user.

\subsection{The \citet{doi:10.1111/j.1538-4616.2007.00014.x} model}

\begin{table}
\begin{centering}
\begin{tabular}{lcccccccccccc}
\hline 
 & CPU & $\max\hat{R}$ &  & \multicolumn{3}{c}{$\sigma$} &  & $z_{t}$ &  & $x_{t}$ &  & $\tau_{t}$\tabularnewline
\cline{5-7} \cline{6-7} \cline{7-7} 
 & time &  &  & Post. & Post. & ESS &  & min &  & min &  & min\tabularnewline
 & (s) &  &  & mean & SD &  &  & ESS &  & ESS &  & ESS\tabularnewline
\hline 
\multirow{2}{*}{EM DR0} & 876 & 1.020 &  & 0.31 & 0.05 & 1825 &  & 782 &  & 189 &  & 916\tabularnewline
 &  &  &  &  &  & {[}2.1{]} &  & {[}0.9{]} &  & {[}0.2{]} &  & {[}1.0{]}\tabularnewline
\hline 
\multirow{2}{*}{EM DR1} & 214 & 1.004 &  & 0.31 & 0.05 & 3950 &  & 2885 &  & 2043 &  & 5317\tabularnewline
 &  &  &  &  &  & {[}18.5{]} &  & {[}13.5{]} &  & {[}9.6{]} &  & {[}24.9{]}\tabularnewline
\hline 
\multirow{2}{*}{Stan DR0} & 86 & 1.061 &  & 0.31 & 0.05 & 842 &  & 186 &  & 143 &  & 387\tabularnewline
 &  &  &  &  &  & {[}9.8{]} &  & {[}2.2{]} &  & {[}1.7{]} &  & {[}4.5{]}\tabularnewline
\hline 
\multirow{2}{*}{Stan DR1} & 30 & 1.038 &  & 0.32 & 0.05 & 1267 &  & 199 &  & 624 &  & 851\tabularnewline
 &  &  &  &  &  & {[}42.1{]} &  & {[}6.6{]} &  & {[}20.7{]} &  & {[}28.3{]}\tabularnewline
\hline 
\multirow{2}{*}{RM} & 658 & 1.007 &  & 0.31 & 0.05 & 1917 &  & 977 &  & 2054 &  & 3273\tabularnewline
 &  &  &  &  &  & {[}2.9{]} &  & {[}1.5{]} &  & {[}3.1{]} &  & {[}5.0{]}\tabularnewline
\hline 
\end{tabular}\caption{\label{tab:Effective-sample-sizes-SW}Effective sample sizes and diagnostics
for the \citet{doi:10.1111/j.1538-4616.2007.00014.x} model (\ref{eq:SW_z}-\ref{eq:SW_y}).
Two variants of Dynamic Rescaling (DR0 and DR1) was applied for the
EM sampler \citep[see][Section 6 for details]{doi:10.1080/10618600.2019.1584901}. }
\par\end{centering}
\end{table}
The final smaller example model considered is the \citet{doi:10.1111/j.1538-4616.2007.00014.x}
quarterly inflation rate model. The model may be summarized by a pair
of latent stochastic volatility processes with first order Gaussian
random walk structure
\begin{align}
z_{t}|z_{t-1},\sigma & \sim N(z_{t-1},\sigma^{2}),\;t=2,\dots,T-1,\label{eq:SW_z}\\
x_{t}|x_{t-1},\sigma & \sim N(x_{t-1},\sigma^{2}),\;t=2,\dots,T.\label{eq:SW_x}
\end{align}
Further, a latent stochastic trend process is modeled as a first order
random walk with stochastic volatility
\begin{equation}
\tau_{t}|\tau_{t-1},z_{t-1}\sim N(\tau_{t-1},\exp(z_{t-1})),\;t=2,\dots,T.\label{eq:SW_tau}
\end{equation}
Finally, the observed time series of inflation rates $\mathbf{y}$
is modeled as
\begin{equation}
y_{t}|\tau_{t},x_{t}\sim N(\tau_{t},\exp(x_{t})),\;t=1,\dots,T.\label{eq:SW_y}
\end{equation}
The model is completed by the prior $\sigma^{-2}\sim$ Gamma(5.0,0.5),
and is applied to the same data set as in \citet{doi:10.1080/10618600.2019.1584901},
namely quarterly log-returns $\times100$ of the US CPI between 1955Q1
and 2018Q1. It is seen that the model involves two layers ($\mathbf{z}$
and $(\mathbf{x},\boldsymbol{\tau})$) of non-linearly coupled latent
variables, which poses substantial challenges for most MCMC methods. 

The model was implemented for the RM sampler with $\mathbf{q}=(\mathbf{z},\mathbf{x},\boldsymbol{\tau},\log(\sigma^{-2}))$
to obtain a tri-diagonal sparsity structure suitable for the sparse
Cholesky factorization employed. As benchmarks, EM- and Stan samplers
based on two modes of Dynamic Rescaling denoted DR0 and DR1 \citep[see][Section 6 for details]{doi:10.1080/10618600.2019.1584901}
were considered. Direct EM or Stan sampling (i.e. with $\mathbf{q}=(\mathbf{z},\mathbf{x},\boldsymbol{\tau},\log(\sigma^{-2}))$)
was not competitive.

Diagnostic results are provided in Table \ref{tab:Effective-sample-sizes-SW}.
It is seen that the proposed methodology produces reliable results
with minimal requirements of the user. Implemented both with EM and
Stan, the DR1 method is more efficient than RM, but it is worth noticing
that the implementation of the DR1 methodology in this case requires
substantial user input- and expertise (essentially involving integrating
out the complete $\boldsymbol{\tau}|\mathbf{z},\mathbf{x},\mathbf{y},\sigma$
using bespoke tri-diagonal Cholesky algorithms). Further, when implemented
in Stan, DR0 has performance roughly on par with RM, but again the
DR0 is also here highly non-trivial to implement.

\section{A Wishart transition random walk stochastic volatility model\label{sec:A-Wishart-transition_SV}}

This section considers a restricted case of the multivariate stochastic
volatility model of \citet{doi:10.1198/073500105000000306}, where
the precision matrix of the log-return vectors follows a random walk
model with Wishart distributed transitions. Denote by $\mathcal{W}_{p}(\mathbf{V},\nu)$
the Wishart distribution on $p\times p$ SPD matrices the for $\nu>p-1$
degrees of freedom and with SPD scale matrix $\mathbf{V}$ (so that
$E(\mathbf{P})=\nu\mathbf{V}$ when $\mathbf{P}\sim\mathcal{W}_{p}(\mathbf{V},\nu)$).
Then the model considered here may be summarized by 
\begin{align}
\mathbf{P}_{t}|\mathbf{P}_{t-1},\nu & \sim\mathcal{W}_{p}(\nu^{-1}\mathbf{P}_{t-1},\nu),\;t=2,\dots,T\label{eq:WSV_1}\\
\mathbf{y}_{t}|\mathbf{P}_{t} & \sim N(\mathbf{0},\mathbf{P}_{t}^{-1}),\;t=1,\dots,T\label{eq:WSV_2}
\end{align}
where $\mathbf{y}_{t}\in\mathbb{R}^{p},\;t=1,\dots,T$, are log-return
vectors of $p$ assets. The model is finalized with the prior $\nu\sim N(250.0,20.0^{2})$,
with no special attention given to the constraint on $\nu$ as $\nu\approx p-1$
is highly unlikely under the posterior distribution considered here.
The data set (with $p=3$ and $T=1095$) under consideration consist
of daily observations of exchange rates of Australian Dollars (AUD),
Canadian Dollars (CAD) and Swiss Francs (CHF) against the US Dollar
between Jan. 2nd 2008 and Apr. 4th 2012. The data are a subset of
the \texttt{exrates} data set from the R package \texttt{stochvol}
\citep{JSSv069i05}.

\subsection{LGCs of SPD matrix-variate distributions}

The $p\times p$ SPD matrices $\mathbf{P}_{t},\;t=1,\dots,T$, are
represented in terms of of unrestricted vectors $\mathbf{z}_{t}\in\mathbb{R}^{p(p+1)/2}$
via the transformation 
\[
\mathbf{P}_{t}=\mathcal{P}(\mathbf{z}_{t})=\mathbf{L}(\mathbf{z}_{t})\boldsymbol{\Lambda}(\mathbf{z}_{t})\mathbf{L}^{T}(\mathbf{z}_{t}),\;\boldsymbol{\Lambda}(\mathbf{z}_{t})=\text{diag}(\exp([\mathbf{z}_{t}]_{1}),\dots,\exp([\mathbf{z}_{t}]_{p})),
\]
and where $\mathbf{L}(\mathbf{z}_{t})$ is unit lower triangular with
the below diagonal columns filled with $[\mathbf{z}_{t}]_{\mathbf{L}}=[\mathbf{z}_{t}]_{p+1:p(p+1)/2}$.
See Appendix \ref{sec:LGCs-and-distributions-P-rep} for details.
Hence internally, the sampled quantities are $\mathbf{q}=(\mathbf{z}_{1},\mathbf{z}_{2},\dots,\mathbf{z}_{T},\nu)$,
but the details of the representation of SPD matrices is hidden from
the user in the model specification code. 

Appendix \ref{subsec:LGC-of-Multivariate-gauss-prec} provides the
LGC $\mathbb{V}[\mathbf{y}_{t}|\mathbf{z}_{t}]$ consistent with $\mathbf{y}_{t}|\mathbf{z}_{t}\sim N(\mathbf{0},\left[\mathcal{P}(\mathbf{z}_{t})\right]^{-1})$
needed to implement (\ref{eq:WSV_2}). Appendix \ref{subsec:Implied-Wishart-distribution-WRW}
gives the distribution of $\mathbf{z}_{t}|\mathbf{z}_{t-1},\nu$ so
that $\mathcal{P}(\mathbf{z}_{t})\sim\mathcal{W}(\nu^{-1}\mathcal{P}(\mathbf{z}_{t-1}),\nu)$,
which is needed to implement the time dynamics of $\{\mathbf{z}_{t}\}_{t}$
consistent with (\ref{eq:WSV_1}). Two variants of the proposed methodology
are considered for $\mathbf{z}_{t}|\mathbf{z}_{t-1},\nu$ consistent
with (\ref{eq:WSV_1}), corresponding to two different factorizations
in target representation (\ref{eq:target_rep}). In the former, denoted
RM-J, $\pi(\mathbf{z}_{t}|\mathbf{z}_{t-1},\nu)$ is considered a
factor in (\ref{eq:target_rep}), and the LGC $\mathbb{V}[\mathbf{z}_{t}|\mathbf{z}_{t-1},\nu]$
may be found in Appendix \ref{subsec:Implied-Wishart-distribution-WRW}.
In the second factorization, denoted RM-F, $\pi(\mathbf{z}_{t}|\mathbf{z}_{t-1},\nu)$
is further factorized as $\pi([\mathbf{z}_{t}]_{1:p}|\mathbf{z}_{t-1},\nu)\pi([\mathbf{z}_{t}]_{L}|[\mathbf{z}_{t}]_{1:p},\mathbf{z}_{t-1},\nu)$.
Here $[\mathbf{z}_{t}]_{i},\;i=1,\dots,p$ under $\pi([\mathbf{z}_{t}]_{1:p}|\mathbf{z}_{t-1},\nu)$
are independent ExpGamma-distributed (with LGC given in (\ref{eq:expGammaLGC})),
and $\pi([\mathbf{z}_{t}]_{L}|[\mathbf{z}_{t}]_{1:p},\mathbf{z}_{t-1},\nu)$
consist of a sequence of independent multivariate Gaussian distributions
with covariance matrices having (different) $\mathcal{P}$-representations,
whose LGCs are given in Appendix \ref{subsec:LGC-of-Mulitvariate-cov}.
Note that $\{\mathbf{z}_{t}\}_{t}$ is Markovian, which leads to an
arrow-head structure of $\mathbf{G}(\mathbf{q})$ which lends itself
well to efficient sparse Cholesky factorization. Also for the EM and
Stan benchmarks for this model were carried out using $\mathbf{q}=(\mathbf{z}_{1},\mathbf{z}_{2},\dots,\mathbf{z}_{T},\nu)$
as the sampled quantity. 

\subsection{Results}

\begin{table}
\begin{centering}
\begin{tabular}{ccccccccccc}
\hline 
 & CPU & $\max\hat{R}$ &  & \multicolumn{3}{c}{$\nu$} &  & \multicolumn{3}{c}{$\mathbf{z}_{t},\;t=1,\dots,T$}\tabularnewline
\cline{5-7} \cline{6-7} \cline{7-7} \cline{9-11} \cline{10-11} \cline{11-11} 
 & time &  &  & post. & post. & ESS &  & min & median & max\tabularnewline
 & (s) &  &  & mean & SD &  &  & ESS & ESS & ESS\tabularnewline
\hline 
\multirow{2}{*}{RM-J} & 24020 & 1.0036 &  & 256.8 & 16.78 & 3880 &  & 2602 & 12738 & 31225\tabularnewline
 &  &  &  &  &  & {[}0.162{]} &  & {[}0.108{]} & {[}0.530{]} & {[}1.300{]}\tabularnewline
\hline 
\multirow{2}{*}{RM-F} & 27451 & 1.0032 &  & 256.9 & 16.87 & 3874 &  & 2780 & 12694 & 31225\tabularnewline
 &  &  &  &  &  & {[}0.141{]} &  & {[}0.101{]} & {[}0.462{]} & {[}1.137{]}\tabularnewline
\hline 
\multirow{2}{*}{EM} & 21635 & 1.1618 &  & 258.7 & 17.01 & 25 &  & 1727 & 5410 & 7093\tabularnewline
 &  &  &  &  &  & {[}0.001{]} &  & {[}0.080{]} & {[}0.250{]} & {[}0.328{]}\tabularnewline
\hline 
\multirow{2}{*}{Stan} & 58293 & 1.018 &  & 254.7 & 17.21 & 264 &  & 2706 & 5579 & 10898\tabularnewline
 &  &  &  &  &  & {[}0.005{]} &  & {[}0.046{]} & {[}0.096{]} & {[}0.187{]}\tabularnewline
\hline 
\end{tabular}
\par\end{centering}
\caption{\label{tab:Wishart-transition-RWSV}Diagnostic results and posterior
moments of $\nu$ under the Wishart transition RWSV model (\ref{eq:WSV_1},\ref{eq:WSV_2})
applied to exchange rate data (AUS, CAD, CHF against USD). Figures
in square brackets are time-weighted ESS (ESS/second computing time).}
\end{table}
\begin{figure}
\centering{}\includegraphics[scale=0.5]{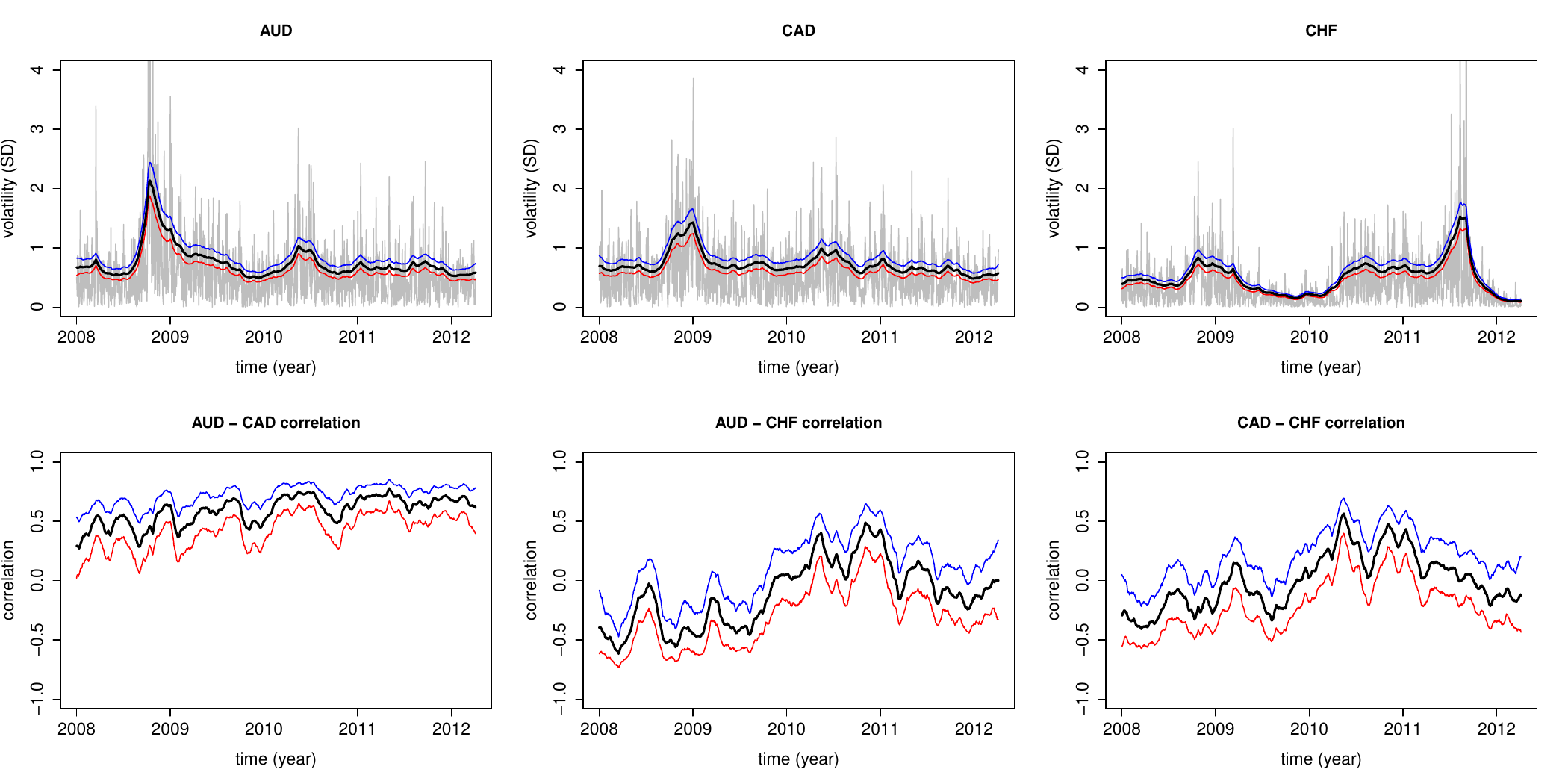}\caption{\label{fig:Marginal-volatilities-and}Marginal volatilities and correlations
associated with the with the Wishart transition RWSV model model (\ref{eq:WSV_1},\ref{eq:WSV_2})
for a single RM-GRHMC trajectory. In the upper panels, the absolute
returns $|y_{i,t}|$ are given in grey, and the lines indicate 0.1,0.5,0.9-quantiles
of the volatility $\sqrt{[\mathbf{P}_{t}^{-1}]_{i,i}}$ (i.e. measured
in standard deviations). In the lower panel, the lines indicate the
0.1,0.5,0.9-quantiles of the correlations implied by the posterior
distributions of each $\mathbf{P}_{t}$.}
\end{figure}
Diagnostic results, and posterior moments of $\nu$ are provided in
Table \ref{tab:Wishart-transition-RWSV}. It is seen that the EM sampler
fails to properly explore the target distribution, whereas both RM
samplers provide reliable results. The Stan sampler uses roughly double
the amount of CPU time, and produces only low ESS for the parameter
$\nu$, which all in all results in that the RM-based methods have
uniformly the best time weighted ESSes.

The difference in raw ESSes between the two RM samplers are rather
small, whereas the RM-F sampling is somewhat slower, leading to slightly
slower time-weighted ESSes. Figure \ref{fig:Marginal-volatilities-and}
presents posterior quantiles of the marginal volatilities and correlations.
It is seen that the model captures substantial time-variation in the
correlations, which is missed in other multivariate SV specifications. 

The MCMC method proposed by \citet{doi:10.1198/073500105000000306}
was a Gibbs sampler involving each updating each $\mathbf{P}_{t}$
using a random walk Metropolis steps, and which would be both time
consuming and require substantial experience to develop. Though a
direct comparison of the (probably highly autocorrelated) output from
a Gibbs sampler against the output of the proposed methodology is
not done here, it is at least clear that the proposed methodology
can produce highly reliable results for large and complicated models
with minimal user intervention. I.e. the specification of the model
requires only a handful of C++ lines corresponding to (\ref{eq:WSV_1},\ref{eq:WSV_2})
and the prior, the rest is handled by software. 

\section{Discussion\label{sec:Discussion}}

Log-gradient covariances and a new metric tensor $\mathbf{G}(\mathbf{q})$
built from log-gradient covariances was proposed. Through numerical
experiment and illustrations, it is shown that the metric tensor in
conjunction with numerical generalized randomized HMC processes allows
pushing the boundary for which hierarchical models can be fitted efficiently.
The methodology is easy to use as the sole responsibility of the user
is to specify the sequence of conditional distribution making up the
model without much regard for imposing special structures in the model.
The nuts and bolts of the proposed methodology, including derivative
calculations and sparse matrix numerical linear algebra may be completely
hidden from the user. 

Deriving LGCs for further models and implementing these in the library
holds scope for further work. Common structures such as linear regression
models, logistic regression models, in addition to Gaussian spatial
models are such examples. Further, adding functionality that hides
the requirement to transform any variable to take values on the complete
real line will also be developed. In addition, this paper only leverages
a subset of what is possible within the NGRHMC framework \citep{kleppe_CTHMC}.
Deriving more adaptive event rates, and corresponding methods for
momentum updates, in the context Riemann manifold NGRHMC is an avenue
that will be pursued.

\bibliographystyle{chicago}
\bibliography{/Users/torekleppe/Dropbox/work/bibtex/kleppe}

\newpage{}

\appendix
\begin{center}
{\LARGE{}Supplementary material to ``Log-density gradient covariance
and automatic metric tensors for Riemann manifold Monte Carlo methods''
by Tore Selland Kleppe}{\LARGE\par}
\par\end{center}

In the following, equations numbers <\ref{eq:num_var_transformation}
refer to equations in the main text.

\section{Details of the numerical implementation\label{sec:Details-of-the-numerical}}

This section provides further details on the implementation of numerical
GRHMC processes found in the pdmphmc package, used in this paper.
The actual simulation of GRHMC processes are done subject to the invertible
(canonical) variable transformation 
\begin{equation}
\mathbf{q}=\mathbf{m}+\mathbf{S}\mathbf{q}^{\prime},\;\mathbf{p}=\mathbf{S}^{-1}\mathbf{p}^{\prime}.\label{eq:num_var_transformation}
\end{equation}
Here, $\mathbf{m}$ should reflect the location/mean of the target
distribution, and $\mathbf{S}$ is a diagonal matrix where the diagonal
elements of $\mathbf{S}$ should reflect the scale of each element
in $\mathbf{q}$ under the target distribution. The transformation
is done for the purpose of obtaining well-scaled Hamilton's equations
suitable for numerical ODE solvers and has the added benefit letting
the unit of time have roughly the same interpretation across models.
The Hamiltonians used for between event dynamics are
\[
\mathcal{H}(\mathbf{q}^{\prime},\mathbf{p}^{\prime})=-\log\bar{\pi}(\mathbf{m}+\mathbf{S}\mathbf{q}^{\prime})+\frac{1}{2}\log(|\bar{\mathbf{G}}(\mathbf{q}^{\prime})|)+\frac{1}{2}\mathbf{p}^{\prime T}\left[\bar{\mathbf{G}}(\mathbf{q}^{\prime})\right]^{-1}\mathbf{p}^{\prime},
\]
where 
\[
\bar{\mathbf{G}}(\mathbf{q}^{\prime})=\mathbf{S}\mathbf{G}(\mathbf{m}+\mathbf{S}\mathbf{q}^{\prime})\mathbf{S},
\]
for the Riemann manifold variant with $\mathbf{G}(\mathbf{q})$ being
the proposed metric tensor derived in the original parameterization,
and simply $\bar{\mathbf{G}}(\mathbf{q}^{\prime})=\mathbf{I}_{D}$
in the fixed metric case. After having obtained samples $\{\mathbf{q}_{i}^{\prime}\}_{i}$
targeting $p(\mathbf{q}^{\prime})\propto\bar{\pi}(\mathbf{m}+\mathbf{S}\mathbf{q}^{\prime})$,
samples targeting the original target distribution $\pi(\mathbf{q})$
are obtained by simply applying the former equation of (\ref{eq:num_var_transformation})
to each of $\{\mathbf{q}_{i}^{\prime}\}_{i}$.

Both types of processes are implemented using the order 5(4) pair
of Runge-Kutta methods developed by \citet{DORMAND198019} and are
subject to a PI type error controller \citep[see e.g. ][ Chapter 17.2]{numrecipes2007}
with both absolute- and relative error tolerances set to $10^{-4}$.
The location $\mathbf{m}$ and diagonal elements of scale $\mathbf{S}$
are set equal to time-integrated \citep[see][]{kleppe_CTHMC} estimates
of the mean and marginal standard deviations of $\mathbf{q}$ respectively.
These estimates are found during the warmup phase of the simulation.
Further, the event intensity $\lambda$ is tuned during warmup using
the no-U-turn approach described in \citet{kleppe_CTHMC}.

\section{Proofs}

\subsection{Proof of Proposition 1\label{subsec:Proof-of-Proposition1}}

Under Assumption 1, clearly $\lim_{x_{i}\rightarrow\infty}\pi(\mathbf{x}|\boldsymbol{\theta})=\lim_{x_{i}\rightarrow-\infty}\pi(\mathbf{x}|\boldsymbol{\theta})=0\;\forall\;i=1,\dots,d$
since $\int\pi(\mathbf{x}|\boldsymbol{\theta})dx_{i}<\infty$.

Part $I$: $\underset{\pi(\mathbf{x}|\boldsymbol{\theta})}{E}\left[\nabla_{\boldsymbol{\theta}}\log\pi(\mathbf{x}|\boldsymbol{\theta})\right]=\mathbf{0}_{p}$
follows directly from conventional likelihood theory \citep[see e.g.][]{pawi:2001}.
For the gradient with respect to $\mathbf{x}$, under Assumption 1,
it is clear that for $i\in[1,\dots,d]$: 
\begin{align*}
\underset{\pi(\mathbf{x}|\boldsymbol{\theta})}{E}\left[\frac{\partial}{\partial x_{i}}\log\pi(\mathbf{x}|\boldsymbol{\theta})\right] & =\int\int\left[\frac{\partial}{\partial x_{i}}\log\pi(\mathbf{x}|\boldsymbol{\theta})\right]\pi(\mathbf{x}|\boldsymbol{\theta})dx_{i}d\mathbf{x}_{-i},\\
 & =\int\int\frac{\partial}{\partial x_{i}}\pi(\mathbf{x}|\boldsymbol{\theta})dx_{i}d\mathbf{x}_{-i},\\
 & =\int\left[\pi(\mathbf{x}|\boldsymbol{\theta})\right]_{x_{i}=-\infty}^{x_{i}=\infty}d\mathbf{x}_{-i},\\
 & =0
\end{align*}
as the density vanishes when $|x_{i}|\rightarrow\infty$. This completes
the proof of part $I$.

Part $II$: $\underset{\pi}{E}\left[-\nabla_{\boldsymbol{\theta}}^{2}\log\pi(\mathbf{x}|\boldsymbol{\theta})\right]=\mathcal{F}(\boldsymbol{\theta})$
follows from conventional likelihood theory. It remains to show that
\[
\mathcal{V}_{i,j}=\underset{\pi}{Cov}\left[\frac{\partial}{\partial x_{i}}\log\pi(\mathbf{x}|\boldsymbol{\theta}),\frac{\partial}{\partial x_{j}}\log\pi(\mathbf{x}|\boldsymbol{\theta})\right]=\underset{\pi}{E}\left[-\frac{\partial^{2}}{\partial x_{i}\partial x_{j}}\log\pi(\mathbf{x}|\boldsymbol{\theta})\right],\;\forall\;i,j\in[1,\dots,d],
\]
 and that
\[
\mathcal{W}_{i,j}=\underset{\pi}{Cov}\left[\frac{\partial}{\partial x_{i}}\log\pi(\mathbf{x}|\boldsymbol{\theta}),\frac{\partial}{\partial\theta_{j}}\log\pi(\mathbf{x}|\boldsymbol{\theta})\right]=\underset{\pi}{E}\left[-\frac{\partial^{2}}{\partial x_{i}\partial\theta_{j}}\log\pi(\mathbf{x}|\boldsymbol{\theta})\right],\;\forall\;i\in[1,\dots,d],\;j\in[1,\dots,p]
\]
In light of part I,
\[
\mathcal{V}_{i,j}=\int\left[\frac{\partial}{\partial x_{i}}\log\pi(\mathbf{x}|\boldsymbol{\theta})\right]\left[\frac{\partial}{\partial x_{j}}\log\pi(\mathbf{x}|\boldsymbol{\theta})\right]\pi(\mathbf{x}|\boldsymbol{\theta})d\mathbf{x},
\]
and 
\[
\mathcal{W}_{i,j}=\int\left[\frac{\partial}{\partial x_{i}}\log\pi(\mathbf{x}|\boldsymbol{\theta})\right]\left[\frac{\partial}{\partial\theta_{j}}\log\pi(\mathbf{x}|\boldsymbol{\theta})\right]\pi(\mathbf{x}|\boldsymbol{\theta})d\mathbf{x}.
\]
Further we have 
\begin{align*}
\underset{\pi}{E}\left[-\frac{\partial^{2}}{\partial x_{i}\partial x_{j}}\log\pi(\mathbf{x}|\boldsymbol{\theta})\right]= & -\underbrace{\int\left[\frac{\partial^{2}}{\partial x_{i}\partial x_{j}}\pi(\mathbf{x}|\boldsymbol{\theta})\right]d\mathbf{x}}_{=A}\\
 & \underbrace{+\int\left[\frac{\partial}{\partial x_{i}}\log\pi(\mathbf{x}|\boldsymbol{\theta})\right]\left[\frac{\partial}{\partial x_{j}}\log\pi(\mathbf{x}|\boldsymbol{\theta})\right]\pi(\mathbf{x}|\boldsymbol{\theta})d\mathbf{x}}_{=\mathcal{V}_{i,j}}
\end{align*}
and 
\begin{align*}
\underset{\pi}{E}\left[-\frac{\partial^{2}}{\partial x_{i}\partial\theta_{j}}\log\pi(\mathbf{x}|\boldsymbol{\theta})\right]= & -\underbrace{\int\frac{\partial^{2}}{\partial x_{i}\partial\theta_{j}}\log\pi(\mathbf{x}|\boldsymbol{\theta})d\mathbf{x}}_{=B}\\
 & +\underbrace{\int\left[\frac{\partial}{\partial x_{i}}\log\pi(\mathbf{x}|\boldsymbol{\theta})\right]\left[\frac{\partial}{\partial\theta_{j}}\log\pi(\mathbf{x}|\boldsymbol{\theta})\right]\pi(\mathbf{x}|\boldsymbol{\theta})d\mathbf{x}}_{=\mathcal{W}_{i,j}}.
\end{align*}
Then it remains to show that both $A=0$ and $B=0$ under Assumption
1:
\begin{align*}
A & =\int\int\left[\frac{\partial^{2}}{\partial x_{i}\partial x_{j}}\pi(\mathbf{x}|\boldsymbol{\theta})\right]dx_{i}d\mathbf{x}_{-i}\\
 & =\int\frac{\partial}{\partial x_{j}}\underbrace{\left[\int\frac{\partial}{\partial x_{i}}\pi(\mathbf{x}|\boldsymbol{\theta})dx_{i}\right]}_{=0\;(\text{see proof of part }I)}d\mathbf{x}_{-i}=0
\end{align*}
\begin{align*}
B= & \int\int\frac{\partial^{2}}{\partial x_{i}\partial\theta_{j}}\log\pi(\mathbf{x}|\boldsymbol{\theta})dx_{i}d\mathbf{x}_{-i}\\
= & \int\frac{\partial}{\partial\theta_{j}}\underbrace{\left[\int\frac{\partial}{\partial x_{i}}\pi(\mathbf{x}|\boldsymbol{\theta})dx_{i}\right]}_{=0\;(\text{see proof of part }I)}d\mathbf{x}_{-i}=0
\end{align*}
This completes the proof.

\subsection{Proof of (\ref{eq:repar_LCG})\label{subsec:Proof-of-()}}

Suppose the original variables/parameterization $\mathbf{v}=(\mathbf{x},\boldsymbol{\theta})$
has density $\pi(\mathbf{x}|\boldsymbol{\theta})$. Define $\boldsymbol{g}_{\boldsymbol{\theta}}(\mathbf{x})=\nabla_{\mathbf{v}}\log\pi(\mathbf{x}|\boldsymbol{\theta})$
so that by the definition of LGC, $\mathbb{V}_{\pi}[\mathbf{x}|\boldsymbol{\theta}](\boldsymbol{\theta})=\underset{\pi(\mathbf{x}|\boldsymbol{\theta})}{Var}(\mathbf{g}_{\boldsymbol{\theta}}(\mathbf{x}))$.
Further, denote by $p(\mathbf{z}|\boldsymbol{\eta})=\pi(\mathbf{a}(\boldsymbol{\eta})+\mathbf{B}\mathbf{z}|\boldsymbol{\Psi}(\boldsymbol{\eta}))|\mathbf{B}|$
and $\mathbf{w}=(\mathbf{z},\boldsymbol{\eta})$. Then, by the chain
rule, 
\[
\nabla_{\mathbf{w}}\log p(\mathbf{z}|\boldsymbol{\eta})=\mathbf{U}^{T}(\boldsymbol{\eta})\mathbf{g_{\boldsymbol{\Psi}(\boldsymbol{\eta})}}(\mathbf{a}(\boldsymbol{\eta})+\mathbf{B}\mathbf{z}),
\]
and finally
\begin{align*}
\mathbb{V}_{p}[\mathbf{z}|\boldsymbol{\eta}] & =\underset{p(\mathbf{z}|\boldsymbol{\eta})}{Var}\left[\nabla_{\mathbf{w}}\log p(\mathbf{z}|\boldsymbol{\eta})\right]=\mathbf{U}^{T}(\boldsymbol{\eta})\underset{p(\mathbf{z}|\boldsymbol{\eta})}{Var}\left[\mathbf{g_{\boldsymbol{\Psi}(\boldsymbol{\eta})}}(\mathbf{a}(\boldsymbol{\eta})+\mathbf{B}\mathbf{z})\right]\mathbf{U}(\boldsymbol{\eta})\\
 & =\mathbf{U}^{T}(\boldsymbol{\eta})\underset{\pi(\mathbf{x}|\boldsymbol{\theta}=\boldsymbol{\Psi}(\boldsymbol{\eta}))}{Var}\left[\mathbf{g}_{\boldsymbol{\Psi}(\boldsymbol{\eta})}(\mathbf{x})\right]\mathbf{U}(\boldsymbol{\eta})=\mathbf{U}^{T}(\boldsymbol{\eta})\left[\mathbb{V}_{\pi}[\mathbf{x}|\boldsymbol{\theta}](\boldsymbol{\Psi}(\boldsymbol{\eta}))\right]\mathbf{U}(\boldsymbol{\eta}).
\end{align*}

\section{Details for the example models}

\subsection{Details for Poisson regression\label{subsec:Details-for-Poisson}}

The Fisher information matrix for $(\eta,g)$ (i.e. dropping the $i$-subscript)
associated with (\ref{eq:ZIPoisson_1},\ref{eq:ZIPoisson_2}) is given
by
\[
\mathcal{F}=\left[\begin{array}{cc}
\exp(\eta)\frac{1+\exp(g+\exp(\eta))-\exp(g+\eta)}{(1+\exp(g))\left[1+\exp(g+\exp(\eta))\right]} & -\frac{\exp(g+\eta+\exp(\eta))}{(1+\exp(g))\left[\exp(g)+\exp(-\exp(\eta))\right]}\\
-\frac{\exp(g+\eta+\exp(\eta))}{(1+\exp(g))\left[\exp(g)+\exp(-\exp(\eta))\right]} & \exp(2g)\frac{\exp(\exp(\eta))-1}{(1+\exp(g))^{2}\left[1+\exp(g+\exp(\eta))\right]}
\end{array}\right].
\]
The responses $\mathbf{y}$ are the number of salamanders observed
(``count'' in data set) The model uses the variable species (``spp'',
which originally has 8 levels, and has been converted to an intercept
term and 7 dummy variables) as the fixed effect-part of both $\boldsymbol{\eta}$
and $\mathbf{g}$. In addition, the random effects in $\boldsymbol{\eta}$
are specific to one of 23 sampling sites (``site'') so that each
$\eta_{i}$ depends additively on a single random effect. The random
effects (conditionally on $\sigma$) have independent $N(0,\sigma^{2})$
priors, and $\sigma^{2}$ has an exponential prior with expectation
1. The call to fit the same model under a frequentist framework (i.e.
with no prior on $\sigma^{2}$) using the glmmTMB package is: \texttt{glmmTMB::glmmTMB(count\textasciitilde spp
+ (1|site),zi=\textasciitilde spp,data=Salamanders,family = poisson)}.

\subsection{Details related to the CEV model with additive noise\label{subsec:Details-related-to-CEV}}

The priors used are $\alpha\sim N(0,100\Delta^{-2})$, $\beta\sim N(\Delta^{-1},100\Delta^{-2})$
where $\Delta=1/252$. Further, flat priors on $\mathbb{R}$ were
used for $\log(\sigma_{x}^{2})$ and $\log(\sigma_{y}^{2})$. Finally,
a flat prior on $(0,\infty)$ was used for $\gamma$.

\section{LGCs and distributions related to the $\mathcal{P}$-representation
of SPD matrices\label{sec:LGCs-and-distributions-P-rep}}

\subsection{The $\mathcal{P}$-representation of SPD matrices\label{subsec:The-P-representation-of}}

Before discussing LGC and distributions related to the $\mathcal{P}$-representation
of SPD matrices, recall that 
\[
\mathcal{P}(\boldsymbol{\omega})=\mathbf{L}(\boldsymbol{\omega})\boldsymbol{\Lambda}(\boldsymbol{\omega})\mathbf{L}^{T}(\boldsymbol{\omega})\in\mathbb{R}^{n\times n},
\]
where $\boldsymbol{\omega}\in\mathbb{R}^{n(n+1)/2}$, 
\[
\boldsymbol{\Lambda}(\boldsymbol{\omega})=\text{diag}(\exp(\omega_{1}),\dots,\exp(\omega_{n}))
\]
and 
\[
\mathbf{L}(\mathbf{x})=\left[\begin{array}{ccccc}
1 & 0 & \cdots & 0 & 0\\
\omega_{\kappa_{1}} & 1 & \cdots & 0 & 0\\
\omega_{\kappa_{1}+1} & \omega_{\kappa_{2}} & \ddots & 0 & 0\\
\vdots & \vdots & \ddots & \vdots & \vdots\\
\omega_{\kappa_{1}+n-3} & \omega_{\kappa_{2}+n-4} & \cdots & 1 & 0\\
\omega_{\kappa_{1}+n-2} & \omega_{\text{\ensuremath{\kappa_{2}+n-3}}} & \cdots & \omega_{\kappa_{n-1}} & 1
\end{array}\right],\;\kappa_{j}=nj-\frac{j(j-1)}{2}+1,j=1,\dots,n-1.
\]
Note that $\kappa_{1}=n+1$ and therefore the elements of $\boldsymbol{\omega}$
appear in \emph{either} $\boldsymbol{\Lambda}$ or $\mathbf{L}$,
and the notation $\boldsymbol{\omega}_{\Lambda}=\boldsymbol{\omega}_{1:n}$
and $\boldsymbol{\omega}_{L}=\boldsymbol{\omega}_{n+1:n(n+1)/2}$
will be used subsequently. Clearly $\mathcal{P}(\boldsymbol{\omega})$
is SPD for any $\boldsymbol{\omega}\in\mathbb{R}^{n(n+1)/2}$. 

Further, define $\mathcal{P}_{r}(\boldsymbol{\omega})=\left[\mathbf{L}(\boldsymbol{\omega})\right]_{r+1:n,r+1:n}\left[\boldsymbol{\Lambda}(\boldsymbol{\omega})\right]_{r+1:n,r+1:n}\left[\mathbf{L}(\boldsymbol{\omega})\right]_{r+1:n,r+1:n}^{T}\in\mathbb{R}^{n-r\times n-r}$
so that $\mathcal{P}_{0}=\mathcal{P}$. Note that $\mathcal{P}_{r}(\boldsymbol{\omega})$
is generally not equal to $\left[\mathcal{P}(\boldsymbol{\omega})\right]_{r+1:n,r+1:n}$.
Further note that $\mathcal{P}_{r}(\boldsymbol{\omega}),\;r>0$ is
it self a lower-dimensional $\mathcal{P}$-representation of a subset
of the elements in $\boldsymbol{\omega}$, i.e. 
\begin{equation}
\mathcal{P}_{r}(\boldsymbol{\omega})=\mathcal{P}(\mathcal{\omega}^{\prime}),\;\text{ where }\boldsymbol{\omega}^{\prime}=(\boldsymbol{\omega}_{r+1:n},\boldsymbol{\omega}_{\kappa_{r+1}:n(n+1)/2}).\label{eq:P-rep-lower-dim}
\end{equation}

As will be clear later, it is often required to explicitly compute
the inverse of $\mathcal{P}(\boldsymbol{\omega})$ and also the inverse
of each of $\mathcal{P}_{r}(\boldsymbol{\omega}),r=n-1,n-2,\dots,1$.
Fortunately, as discussed in the supplementary material of \citet{doi:10.1080/10618600.2019.1584901},
these inverses may be computed rather easily using the recursion
\[
\mathcal{P}_{n-1}^{-1}(\boldsymbol{\omega})=\left[\exp(-\omega_{n})\right],\;\mathcal{P}_{r}^{-1}(\boldsymbol{\omega})=\left[\begin{array}{cc}
\exp(-\omega_{r+1})+\boldsymbol{\rho}_{r+1}^{T}\boldsymbol{\omega}_{J_{r+1}} & -\boldsymbol{\rho}_{r+1}^{T}\\
-\boldsymbol{\rho}_{r+1} & \mathcal{P}_{r+1}^{-1}(\boldsymbol{\omega})
\end{array}\right],r=n-2,n-3,\dots,0
\]
where $J_{j}=\kappa_{j}:(\kappa_{j+1}-1)$ and $\boldsymbol{\rho}_{r}=\left[\mathcal{P}_{r}^{-1}(\boldsymbol{\omega})\right]\boldsymbol{\omega}_{J_{r}}$.

A further convenient fact is that the mapping between $\boldsymbol{\omega}$
and the elements in either the upper- or lower triangular part of
$\mathcal{P}(\boldsymbol{\omega})$, i.e. the transformation $\boldsymbol{\omega}\mapsto\text{vech}(\mathcal{P}(\boldsymbol{\omega}))$,
is bijective and has Jacobian determinant proportional to $\exp(\sum_{i=1}^{n}(n+1-i)\omega_{i})$. 

\subsection{LGC of Multivariate Gaussian with precision matrix $\alpha\mathcal{P}(\boldsymbol{\omega})$\label{subsec:LGC-of-Multivariate-gauss-prec}}

Let $\alpha>0$ be a scalar. Then the multivariate Gaussian with precision
matrix $\alpha\mathcal{P}(\boldsymbol{\omega})$ and density $\mathcal{N}(\mathbf{x}|\boldsymbol{\mu},[\alpha\mathcal{P}(\boldsymbol{\omega})]^{-1})$
has the LGC \begin{equation*}
\mathbb{V}_{\mathcal{N}}[\mathbf{x}|\boldsymbol{\mu},\alpha,\boldsymbol{\omega}]= 
\begin{blockarray}{cccccc}
 & \mathbf x & \boldsymbol \mu & \alpha & \boldsymbol \omega_\Lambda & \boldsymbol \omega_L \\
\begin{block}{c(ccccc)}
\mathbf x & \alpha\mathcal{P}(\boldsymbol{\omega}) & -\alpha\mathcal{P}(\boldsymbol{\omega}) & \mathbf 0 & \mathbf 0 & \mathbf 0 \\
\boldsymbol \mu & -\alpha\mathcal{P}(\boldsymbol{\omega}) & \alpha\mathcal{P}(\boldsymbol{\omega}) & \mathbf 0 & \mathbf 0 & \mathbf 0 \\
\alpha & \mathbf 0 & \mathbf 0 & \frac{n}{2\alpha^2} & \frac{1}{2\alpha} \mathbf 1^T & \mathbf 0 \\
\boldsymbol \omega_\Lambda & \mathbf 0 & \mathbf 0 & \frac{1}{2\alpha} \mathbf 1 & \frac{1}{2}\mathbf I & \mathbf 0 \\
\boldsymbol \omega_L & \mathbf 0 & \mathbf 0 & \mathbf 0 & \mathbf 0 & \mathcal F_{\boldsymbol \omega_L} \\
\end{block}
\end{blockarray}
\end{equation*}where $\mathcal{F}_{\boldsymbol{\omega}_{L}}=\text{bdiag}[\exp(\omega_{1})\mathcal{P}_{1}^{-1}(\boldsymbol{\omega}),\exp(\omega_{2})\mathcal{P}_{2}^{-1}(\boldsymbol{\omega}),\dots,\exp(\omega_{n-1})\mathcal{P}_{n-1}^{-1}(\boldsymbol{\omega})]$

\subsection{LGC of Mulitvariate Gaussian with covariance matrix $\alpha\mathcal{P}(\mathbf{\omega})$\label{subsec:LGC-of-Mulitvariate-cov}}

The multivariate Gaussian with covariance matrix $\alpha\mathcal{P}(\mathbf{\omega})$
and density $\mathcal{N}(\mathbf{x}|\boldsymbol{\mu},\alpha\mathcal{P}(\mathbf{\omega}))$
has the LGC\begin{equation}
\mathbb{V}_{\mathcal{N}}[\mathbf{x}|\boldsymbol{\mu},\alpha,\boldsymbol{\omega}]= 
\begin{blockarray}{cccccc}
 & \mathbf x & \boldsymbol \mu & \alpha & \boldsymbol \omega_\Lambda & \boldsymbol \omega_L \\
\begin{block}{c(ccccc)} 
\mathbf x & [\alpha\mathcal P(\boldsymbol \omega)]^{-1} & - [\alpha\mathcal P(\boldsymbol \omega)]^{-1} & \mathbf 0 & \mathbf 0 & \mathbf 0 \\
\boldsymbol \mu & -[\alpha\mathcal P(\boldsymbol \omega)]^{-1} &  [\alpha\mathcal P(\boldsymbol \omega)]^{-1} & \mathbf 0 & \mathbf 0 & \mathbf 0 \\
\alpha & \mathbf 0 & \mathbf 0 & \frac{n}{2 \alpha^2} & \frac{1}{2 \alpha} \mathbf 1^T & \mathbf 0 \\
\boldsymbol \omega_\Lambda & \mathbf 0 & \mathbf 0 & \frac{1}{2 \alpha} \mathbf 1 & \frac{1}{2} \mathbf I & \mathbf 0 \\
\boldsymbol \omega_L & \mathbf 0 & \mathbf 0 & \mathbf 0 & \mathbf 0 & \mathcal F_{\boldsymbol \omega_L}\\
\end{block}
\end{blockarray}
\label{eq:mvnorm_var_lgc}
\end{equation}where $\mathcal{F}_{\boldsymbol{\omega}_{L}}$ is given in the previous
section.

\subsection{Implied diagonal scale matrix Wishart distribution}

Consider an $n\times n$ SPD matrix $\mathbf{P}\sim\mathcal{W}(\mathbf{W},\nu)$
where $\mathbf{W}=\text{diag}(w_{1},\dots,w_{n})=\text{diag}(\mathbf{w})$
is diagonal and positive definite, i.e. so that $E(\mathbf{P})=\nu\mathbf{W}$.
Rather than deriving the LGC in the $\mathbf{P}$-representation directly,
the distribution on $\boldsymbol{\omega}$ consistent with $\mathcal{P}(\boldsymbol{\omega})\sim\mathcal{W}(\mathbf{W},\nu)$
may be derived either using the Bartlett representation, or via the
general Wishart density and transformation formula since $\boldsymbol{\omega}\mapsto\text{vech}(\mathcal{P}(\boldsymbol{\omega}))$
is bijective. Either way, we end up with the following hierarchical
representation
\begin{align*}
\omega_{i}|\nu,w_{i} & \sim\text{ExpGamma}(\nu/2-(i-1)/2,2w_{i}),\;i=1,\dots,n,\\
\boldsymbol{\omega}_{J_{k}}|\omega_{k},\mathbf{W} & \sim N\left(\mathbf{0},\exp(-\omega_{k})\mathbf{W}_{k+1:n,k+1:n}\right),\;k=1,\dots,n-1.
\end{align*}
In the present implementation, the LGCs of $\omega_{i}|\nu,w_{i},\;i=1,\dots,n$
(see Equation \ref{eq:expGammaLGC}) and $\boldsymbol{\omega}_{J_{k}}|\omega_{k},\mathbf{W},\;k=1,\dots,n-1$
(see Equation \ref{eq:1dnormal_LGC} as $\boldsymbol{\omega}_{L}|\boldsymbol{\omega}_{\Lambda},\mathbf{W}$
are independent) and combined via (\ref{eq:the_metric_tensor}).

\subsection{Implied Wishart distribution for scale matrix on the form $\nu^{-1}\mathcal{P}(\mathbf{y})$\label{subsec:Implied-Wishart-distribution-WRW}}

The implied distribution of $\mathbf{x}|\mathbf{y},\nu$ consistent
with $\mathcal{P}(\mathbf{x})\sim\mathcal{W}_{d}(\nu^{-1}\mathcal{P}(\mathbf{y}),\nu)$
is given by
\begin{align}
x_{i}|y_{i},\nu & \sim ExpGamma((\nu+1-i)/2,2\nu^{-1}\exp(y_{i})),\;i=1,\dots,n,\label{eq:Iwishart1}\\
\mathbf{x}_{J_{k}}|\mathbf{y},\nu,x_{k} & \sim N(\mathbf{y}_{J_{k}},\nu^{-1}\exp(-x_{k})\mathcal{P}_{k}(\mathbf{y})),\;k=1,\dots,n-1,\label{eq:Iwishart2}
\end{align}
Two variants of LGCs - MC and J were considered for this model in
Section \ref{sec:A-Wishart-transition_SV}. For variant RM-F (factorized),
the LGCs of each of $x_{i}|y_{i},\nu,\;i=1,\dots,n$ (see Equation
\ref{eq:expGammaLGC}) and $\mathbf{x}_{J_{k}}|\mathbf{y},\nu,x_{k},\;k=1,\dots,n-1$
(see Equations \ref{eq:P-rep-lower-dim} and \ref{eq:mvnorm_var_lgc})
are combined via (\ref{eq:the_metric_tensor}). For variant J (joint)
the LGC of the joint distribution of $\mathbf{x}|\mathbf{y},\nu$
obtained by combining (\ref{eq:Iwishart1}) and (\ref{eq:Iwishart2}),
say $\mathbb{V}[\mathbf{x}|\mathbf{y},\nu]$ may be summarized as:
\begin{align*}
\mathbf{x}_{1:n},\mathbf{x}_{1:n} & :\text{diag}(\{\frac{\nu+1+d}{2}-i\}_{i=1}^{n}),\\
\mathbf{x}_{1:n},\mathbf{x}_{J_{k}} & :\mathbf{0},\;k=1,\dots,n-1\\
\mathbf{x}_{i},\mathbf{y}_{j} & :\begin{cases}
-\frac{\nu+1-i}{2} & \text{for }i=j\\
-\frac{1}{2} & \text{for }i<j\\
0 & \text{otherwise}
\end{cases},\;i,j=1,\dots,n\\
\mathbf{x}_{1:n},\mathbf{y}_{J_{k}} & :\mathbf{0},\;k=1,\dots,n-1,\\
\mathbf{x}_{i},\nu & :\frac{n+1-2i}{2\nu},\;i=1,\dots,n,\\
\mathbf{x}_{J_{k}},\mathbf{x}_{J_{k}} & :\exp(-y_{k})(\nu+1-k)^{-1}\mathcal{P}_{k}^{-1}(\mathbf{y}),k=1,\dots,n-1,\\
\mathbf{x}_{J_{k}},\mathbf{x}_{J_{l}} & :\mathbf{0}\;\text{for }k\neq l\\
\mathbf{x}_{J_{k}},\mathbf{y}_{j} & :\mathbf{0}\;\text{for }k=1,\dots,n-1,j=1,\dots,n\\
\mathbf{x}_{J_{k}},\mathbf{y}_{J_{k}} & :-\exp(-y_{k})(\nu+1-k)^{-1}\mathcal{P}_{k}^{-1}(\mathbf{y})\\
\mathbf{x}_{J_{k}},\mathbf{y}_{J_{l}} & :\mathbf{0}\;\text{for }k\neq l\\
\mathbf{x}_{J_{k}},\nu & :\mathbf{0}\;\text{for }k=1,\dots,n-1\\
\mathbf{y}_{1:n},\mathbf{y}_{1:n} & :\frac{\nu}{2}\mathbf{I}_{n}\\
\mathbf{y}_{1:n},\mathbf{y}_{J_{k}} & :\mathbf{0},\;k=1,\dots,n-1,\\
\mathbf{y}_{1:n},\nu & :\mathbf{0}\\
\mathbf{y}_{J_{k}},\mathbf{y}_{J_{k}} & :\exp(-y_{k})\nu{}^{-1}\mathcal{P}_{k}^{-1}(\mathbf{y}),k=1,\dots,n-1\\
\mathbf{y}_{J_{k}},\mathbf{y}_{J_{l}} & :\mathbf{0},\;\text{for }k\neq l,\\
\mathbf{y}_{J_{k}},\nu & :\mathbf{0},\;\text{for }k=1,\dots,n-1\\
\nu,\nu & :-\frac{n}{2\nu}+\text{\ensuremath{\sum_{j=1}^{n}\frac{\Psi^{\prime}(\frac{1}{2}(\nu+1-j))}{4}}}
\end{align*}

\section{Salamander mating model\label{sec:Salamander-mating-model}}

Here a ``crossed'' random effects model for the salamander mating
data of \citet[Chapter 14.5]{mccu:neld:1989} is considered. The model
is the same as the INLA example model ``Salamander model B'' (see
\href{https://sites.google.com/a/r-inla.org/www/examples/volume-ii}{https://sites.google.com/a/r-inla.org/www/examples/volume-ii})
and was also considered by \citet{doi:10.1080/10618600.2019.1584901}.
The dataset contains three ``sub-experiements'' (indexed by $k$),
each involving 20 female ($F$) salamanders (indexed by $i$) and
20 male ($M$) salamanders (indexed by $j$). Random effects specific
to each individual salamander in each experiment have the conditional
priors (latter subscript index is $k$)
\begin{align*}
\left(\begin{array}{c}
b_{i1}^{F}\\
b_{i2}^{F}
\end{array}\right)|\mathbf{P}_{F} & \sim\text{iid }N(\mathbf{0},\mathbf{P}_{F}^{-1}),\;i=1,\dots,20,\;\left(\begin{array}{c}
b_{j1}^{M}\\
b_{j2}^{M}
\end{array}\right)|\mathbf{P}_{M}\sim\text{iid }N(\mathbf{0},\mathbf{P}_{M}^{-1}),\;j=1,\dots,20,\\
b_{i3}^{F}|\tau_{F} & \sim\text{iid }N(0,\tau_{F}^{-1}),\;i=1,\dots,20,\;b_{j3}^{M}|\tau_{M}\sim\text{iid }N(0,\tau_{M}^{-1}),\;j=1,\dots,20.
\end{align*}
Priors for the random effects variance structure are given by $\mathbf{P}_{F},\mathbf{P}_{M}\sim\text{iid }\mathcal{W}(\text{diag}(0.804,0.804),3)$
and $\tau_{F},\tau_{M}\sim\text{iid Gamma}(1,0.622)$. The SPD matrices
$\mathbf{P}_{F},\mathbf{P}_{M}$ are represented using the techniques
of Section \ref{subsec:The-P-representation-of} for both RM and EM
samplers. Note that the random effects across the two first sub-experiments
are allowed to be dependent as these experiments involved the same
salamanders at different points in time. Binary mating outcomes $y_{ijk}$
were recorded for a total of $360$ combinations of female ($i$)
and male ($j$) salamanders across the 3 sub-experiments ($k$), along
with covariates (including an intercept term) $\mathbf{x}_{ijk}\in\mathbb{R}^{5}$.
Finally $P(y_{ijk}=1)$ is modeled as $\text{logit}^{-1}(\mathbf{x}_{ijk}^{T}\boldsymbol{\beta}+b_{ik}^{F}+b_{jk}^{M})$. 

\begin{table}
\centering{}%
\begin{tabular}{lccccc}
 & \multicolumn{2}{c}{RM} &  & \multicolumn{2}{c}{EM}\tabularnewline
\cline{2-3} \cline{3-3} \cline{5-6} \cline{6-6} 
CPU time & \multicolumn{2}{c}{1731 s} &  & \multicolumn{2}{c}{135 s}\tabularnewline
$\max\hat{R}$ & \multicolumn{2}{c}{1.0038} &  & \multicolumn{2}{c}{1.0040}\tabularnewline
\hline 
 & \multicolumn{2}{c}{ESS} &  & \multicolumn{2}{c}{ESS}\tabularnewline
$[\mathbf{P}_{F}]_{1,1}$ & 5161 & {[}3.0{]} &  & 2122 & {[}15.7{]}\tabularnewline
$[\mathbf{P}_{F}]_{1,2}$ & 5455 & {[}3.2{]} &  & 5101 & {[}37.8{]}\tabularnewline
$[\mathbf{P}_{F}]_{2,2}$ & 5219 & {[}3.0{]} &  & 2941 & {[}21.8{]}\tabularnewline
$[\mathbf{P}_{M}]_{1,1}$ & 9111 & {[}5.3{]} &  & 3124 & {[}23.2{]}\tabularnewline
$[\mathbf{P}_{M}]_{1,2}$ & 11915 & {[}6.9{]} &  & 2469 & {[}18.3{]}\tabularnewline
$[\mathbf{P}_{F}]_{2,2}$ & 7953 & {[}4.6{]} &  & 3243 & {[}24.0{]}\tabularnewline
$\tau_{F}$ & 7826 & {[}4.5{]} &  & 2138 & {[}15.9{]}\tabularnewline
$\tau_{M}$ & 6461 & {[}3.7{]} &  & 2889 & {[}21.4{]}\tabularnewline
$\boldsymbol{\beta}$ & $\geq6135$ & {[}$\geq3.5${]} &  & $\geq5236$ & {[}$\geq38.8${]}\tabularnewline
$(b_{ik}^{F},b_{jk}^{M})$ & $\geq5739$ & {[}$\geq3.3${]} &  & $\geq3819$ & {[}$\geq28.3${]}\tabularnewline
\hline 
\end{tabular}\caption{\label{tab:Effective-sample-sizes-salamander-mating}Effective sample
sizes and other diagnostic information for Salamander mating model.}
\end{table}
Table \ref{tab:Effective-sample-sizes-salamander-mating} provides
ESSes, $\hat{R}$s and CPU times for both RM and EM samplers applied
to the above model. It is seen that for this model, the EM sampler
outperforms the RM sampler in terms of sampling efficiency. The very
long computing time for the RM sampler in part stem from the fact
that the sparsity pattern of $\mathbf{G}(\mathbf{q})$ does not lead
to substantial saving since the Cholesky factor is rather dense. Some
improvements of CPU time, and therefore sampling efficiency are conceivable
by permuting the ordering of the random effects, but this is not explored
further here.
\end{document}